\documentclass[twocolumn,aps,prd,showpacs,floatfix]{revtex4}
\usepackage{graphicx}
\usepackage{times}
\usepackage{nicefrac}
\usepackage{amsmath}
\usepackage{amsfonts}
\usepackage{amssymb}
\usepackage{amsthm}
\usepackage{epsf}
\usepackage{bm}
\usepackage{times}

\usepackage{dcolumn}

\newcolumntype{.}{D{x}{}{-1}}

\newcommand{\bfr}{{\bm r}}

\newcommand{\bfn}{{\bm n}}

\newcommand{\be}{\begin{eqnarray}}
\newcommand{\ee}{\end{eqnarray}}

\begin{document}

\title{How to access QED at supercritical Coulomb field}

\author{R.~V.~Popov$^1$, V.~M.~Shabaev$^1$\footnote{Corresponding author: v.shabaev@spbu.ru},
  D.~A.~Telnov$^1$, I.~I.~Tupitsyn$^1$, I.~A.~Maltsev$^1$,
  Y.~S.~Kozhedub$^1$, A.~I.~Bondarev$^2$, N.~V.~Kozin$^1$, X.~Ma$^3$, G.~Plunien$^4$,
T.~St\"ohlker$^{5,6,7}$, D.~A.~Tumakov$^1$, and V.~A.~Zaytsev$^1$}

\affiliation {$^1$Department of Physics, St. Petersburg State University,
Universitetskaya 7/9, 199034 St. Petersburg, Russia\\
$^2$Center for Advanced Studies, Peter the Great St. Petersburg Polytechnic
University, Polytekhnicheskaja 29, 195251 St.Petersburg,  Russia\\
$^3$ Institute of Modern Physics, Chinese Academy of Sciences,
Nanchang rd. 509, 730000 Lanzhou, China\\
$^4$Institut f\"ur Theoretische Physik, TU Dresden,
D-01062 Dresden, Germany\\
$^5$GSI Helmholtzzentrum f\"ur Schwerionenforschung GmbH, D-64291
Darmstadt, Germany\\
$^6$Helmholtz-Institut Jena, D-07743 Jena, Germany\\
$^7$Institut f\"ur Optik und Quantenelektronik,
Friedrich-Schiller-Universit\"at Jena, D-07743 Jena, Germany}

\begin{abstract}
  In slow collisions of two bare nuclei with the total charge number
  larger than the critical value, $Z_{\rm cr} \approx 173$,  the initially neutral vacuum can spontaneously
  decay into the charged vacuum and  two positrons. Detection of the spontaneous emission of
  positrons would be the  direct evidence of this fundamental phenomenon. However,
  the spontaneous emission is generally masked by the dynamical positron emission,
  which is induced by a strong time-dependent electric field created by the colliding nuclei.
  In our  recent paper [I.A. Maltsev et al., Phys. Rev. Lett. {\bf 123}, 113401 (2019)]
  it has been shown that
  the spontaneous pair production  can be observed via
  measurements of the pair-production probabilities for a given set of nuclear trajectories.
  In the present paper, we have significantly advanced this study by exploring additional aspects
  of the process we are interested in. We calculate the positron energy spectra and find
  that these spectra can give a clear signature of
  the transition from the  subcritical to the supercritical regime.
  It is found that focusing on a part of the positron spectrum,
  which accounts for the energy region where the spontaneously created positrons can contribute,
  allows to get a much stronger evidence of the transition
  to the supercritical mode, making it very well pronounced in collisions, for example,
  of two uranium nuclei.   The possibility of extending this study
  to collisions of bare nuclei  with neutral atoms is also considered.
  The probability of a vacancy in the lowest-energy state
  of a quasimolecule which is formed in collisions of a bare U nucleus with neutral
  U and Cm atoms has been calculated. The relatively large values of this probability
  make such collisions suitable for observing the vacuum decay.

\end{abstract}
\maketitle

\section{Introduction}

After the foundations of quantum field theory were
formulated in the early 30s of the last century, it was shown that the theory
predicts the spontaneous creation of electron-positron pairs by a constant
uniform electric field if the strength of the field is comparable to or
greater than a critical value, $E_{\rm cr}=m_e^2c^3/(|e|\hbar) \approx 1.3\times 10^{16}$ V/cm,
\cite{sau31,hei36,sch51}.
While from the point of view of the theory this phenomenon was studied in
a large number of works (see Refs.
\cite{fra91,pia12,gav96,ave02,nar04,sch08,ruf09,abd13,ott15,gav17,tor18,koh18,ale18}
and references therein), its
experimental detection was not possible because of practical inaccessibility of
the required field strength.
Some hopes for experimental detection of this fundamental effect were associated with
the development of novel laser technologies  \cite{rin01,mou06}.
But, apparently, in the foreseeable future, it is quite unlikely
to realize a possible experiment in the required  strong field regime
solely on the basis of high-power laser.

\begin{figure}
\begin{center}
  \includegraphics[width=\columnwidth]{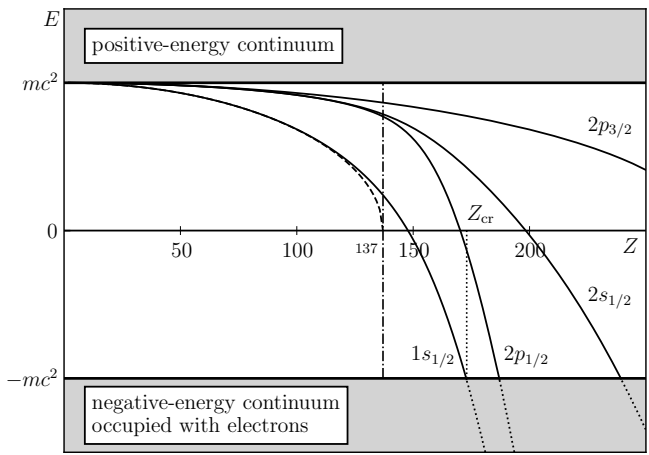}
\end{center}
\caption{The low-lying energy levels of a H-like ion as functions of the nuclear charge number $Z$.}
\label{fig:energy_z}
\end{figure}

\begin{figure}
\begin{center}
  \includegraphics[width=\columnwidth]{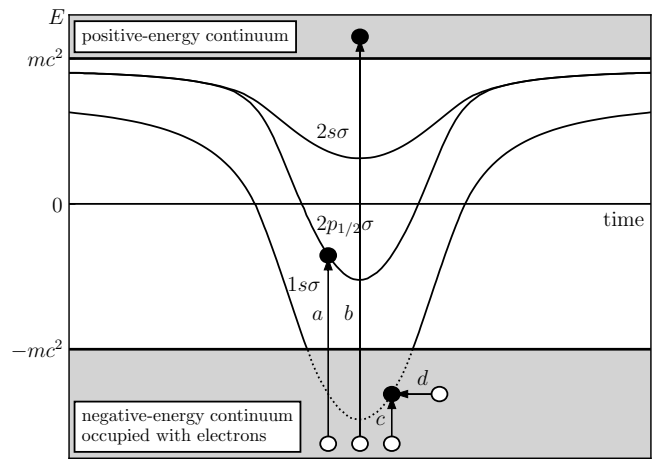}
\end{center}
\caption{The low-lying energy levels of a quasimolecule formed in collision
  of two uranium ions (nuclei)
  as functions of time. The arrows $a$, $b$, and $c$ denote different
  dynamical pair-creation mechanisms
 and the arrow $d$ indicates the  spontaneous pair creation.
 The  $1s$ state dives into the negative-energy continuum for about
 10$^{-21}$ s.
}
\label{fig:energy_mol}
\end{figure}

An alternative approach to the study of vacuum properties in the presence
of a supercritical field was proposed
in the works of Soviet and German physicists
\cite{pom45,ger69,pie69,pop70,zel71,mul72,mur76,pop76,mul76,rei76,sof77,raf78,rei81,gre85,mul88,mul94}.
It is known that
in the case of a pure Coulomb field induced by a point-like charge $Z$
the $1s$ level exists only up to $Z\approx 137$ and then
disappears. However, for extended nuclei this level goes continuously down and
at $Z=Z_{\rm cr}\approx 173$ reaches the onset of the negative-energy Dirac continuum (see
Fig.~\ref{fig:energy_z}).
If this level was empty, it dives into the negative-energy continuum as a resonance,
the decay of which should lead to a spontaneous creation of positrons.
Namely, when switching from the
subcritical to the supercritical mode, the initially neutral vacuum decays into
a charged vacuum and two positrons (due to spin degeneracy).
Since there are no nuclei with such a large
charge in nature (the charge of the heaviest element discovered to date, oganesson, is 118),
the only way to create a supercritical Coulomb field is to collide nuclei with the total charge number
greater than the critical value ($Z_1+Z_2 > 173$). The time dependence of
the  quasi-molecular energy levels in collision of two uranium ions (nuclei) is presented in
 Fig.~\ref{fig:energy_mol}.
Following  Greiner and co-authors \cite{gre85}, in this figure
various pair-creation mechanisms are conventionally shown by arrows.
The arrows $a$, $b$, and $c$ denote dynamical pair-creation
processes and the arrow $d$ indicates the  spontaneous pair creation.
The dynamical pair creation
takes place in both subcritical and supercritical modes while
the spontaneous pair creation starts to work only in the
supercritical regime.
The dynamical mechanisms ($a$, $b$, $c$)
superpose with the  spontaneous one ($d$) and generally can not be
separated from each other. In addition, the spontaneous pair creation is strongly
suppressed due to very small period of the  supercritical regime time ($\sim 10^{-21}$ s),
which is by about two orders of magnitude smaller than the time required for the vacuum
decay.
All this was one of the main reasons why
attempts to experimentally observe the
spontaneous creation of positrons,
which were undertaken more than 30 years ago in Darmstadt (see, e.g., Ref. \cite{gre85} and
references therein), were not successful.
Moreover, the Frankfurt group, which worked on the theory of this phenomenon
for more than 20 years, concluded that the  vacuum decay can
only be observed experimentally
if colliding nuclei stick together for some time due to nuclear forces
\cite{rei05,raf16}.
However, no evidence of  the nuclear sticking in the collisions of interest
has been found to date,
 making this scenario not promising for future experiments.

Despite the aforementioned conclusions by the Frankfurt group, one could
expect, however, that the detailed study of quantum dynamics of the electron-positron field
in low-energy heavy-ion collisions would allow to find some signatures which
indicate the principal difference between the subcritical and
the supercritical regimes.
To carry out these studies, first of all it was necessary to develop the theoretical and
computational methods beyond the approximation made by the Frankfurt group.
To this end,
more than a decade ago new efforts have been initiated by the St. Petersburg group
\cite{tup10,tup12,dey12,dey13,koz14,mal17,pop18,mal18}. These methods allowed
the calculations of the charge-transfer, electron-excitation, ionization, and
pair-production processes. In particular, in Refs. \cite{mal17,pop18,mal18}
the first calculations of the
pair-production probabilities beyond the monopole approximation, which was widely used by the
Frankfurt group, were performed.
These calculations showed that effects beyond
the monopole approximation only slightly change the pair-creation
probabilities in the region of small impact parameters.
This has provided much more flexibility in studying various scenarios
of finding the signatures which can demonstrate
the principal difference between the subcritical and the supercritical regimes.
As a result of these studies, in Ref. \cite{mal19} it was found that
the vacuum decay can be observed via impact-sensitive measurements of
pair-production probabilities. In this paper, we present  additional
aspects of this study and find  important signatures
of the principal difference between the subcritical and supercritical regimes,
which can be observed in the positron spectra.

In the next sections we will mainly consider quantum dynamics of electron-positron
field in collisions of heavy bare nuclei. Experiments on low-energy collisions
of bare nuclei will be possible at the future facilities in
Germany (GSI/FAIR) \cite{gum09,les16,hagXX}, China (HIAF) \cite{ma17}, and
Russia (NICA) \cite{ter15}. After the formulation of the basic
 theory, we will present the results of the calculations of the pair-creation
probabilities and positron spectra as functions of the nuclear charges and
the collision energy for the trajectrories with a given minimal internuclear
distance. The dependence of the pair-creation probabilies and the positron spectra
on the value of the minimal internuclear distance will be also studied.
It will be shown that the study of a partial probability, which accounts
for the region of the positron energy distribution where the spontaneously created
positrons can contribute, allows a significant increase of the signature
of the transition to the supercritical regime, compared to the analysis of
the total probability.
At the end of the paper we will discuss
a possibility for studying the proposed scenarios on the vacuum decay
observation  in collisions of bare nuclei with neutral atoms.

The relativistic units ($\hbar=c=1$) and the Heaviside charge unit
($\alpha=e^2/(4\pi)$, $e<0$) are used throughout the paper.

%

\section{Theory}

\subsection{General formalism}
To study the quantum dynamics of the electron-positron field in low-energy
collisions of heavy nuclei, one can use the approach, where the relative
motion of the colliding nuclei is treated classically \cite{mul72,gre85}.
In this approach, to derive the pair-creation probabilities, one has
to consider first the solutions of  the Dirac equation in the external
time-dependent potential induced
by the colliding nuclei, whose motion is described by the  Rutherford
 trajectories.
 For the relatively slow collisions of interest the magnetic part of the potential can be
 neglected and the time-dependent Dirac equation (TDDE) takes
 the form:
\begin{align}
     \label{eqn:psi_t}
     i\frac{\partial}{\partial t} \psi (\bm r, t) = H(t) \psi (\bm r, t)
\end{align}
with
\begin{align}
 H(t) = \bm{\alpha}\cdot\bm{p} + \beta m_e  + V\left(\bm r, t \right).
\end{align}
Here $\bm{\alpha}$, $\beta$ are the Dirac matrices,
$m_e$ is the electron mass, and $V(\bfr,t)$ is the total two-center potential induced by the nuclei:
\begin{align} \label{pot2}
 V\left(\bm r, t \right) =
 V_{\rm A} \left(|\bm r - \bm R_{\rm A}(t)| \right) +
 V_{\rm B} (|\bm r - \bm R_{\rm B}(t)|),
\end{align}
where the vectors $\bm R_{\rm A}$ and $\bm R_{\rm B}$ denote
the positions of the colliding nuclei and
\begin{equation}
 V_{\rm \, A,B}(\bfr) = \frac{e}{4\pi}
 \int d \bm r^\prime \,
 \frac{\rho_{\rm \, A, B} (\bm r^\prime)}{|\bm r - \bm r^\prime|}
\end{equation}
are the corresponding nuclear potentials.
For the nuclear charge distribution $\rho (\bm r)$ we utilize the model of the uniformly charged sphere.

Considering the quantum dynamics from the initial time $t_{\rm in}$ to the final time  $t_{\rm out}$, one can define
two sets of solutions of the TDDE, which have  the asymptotics:
\begin{eqnarray} \label{init_cond}
  \psi_i^{(+)}(\bfr, t_{\rm in})=\phi_i^{\rm in}(\bfr)\,,\;\;\;\;\;\;
   \psi_i^{(-)}(\bfr, t_{\rm out})=\phi_i^{\rm out}(\bfr)\,,
\end{eqnarray}
where $\phi_i^{\rm in}(\bfr)$ and $\phi_i^{\rm out}(\bfr)$ are the eigenfunctions of the Dirac
Hamiltonian at the corresponding time moments,
\begin{eqnarray}
 H(t_{\rm in})\phi_i^{\rm in}(\bfr)&=&
\varepsilon_i^{\rm in}\phi_i^{\rm in}(\bfr)\,,\\
 H(t_{\rm out})\phi_i^{\rm out}(\bfr)&=&
\varepsilon_i^{\rm out}\phi_i^{\rm out}(\bfr)\,.
\end{eqnarray}
The most natural way to derive formulas for the pair-creation probabilities is to use
the second quantization formalism \cite{fra91}. To this end, we introduce the ``in'' and
``out'' vacuum states, $|0,{\rm in}\rangle$ and $|0,{\rm out}\rangle$,  and the related
annihilation operators:
\begin{eqnarray}
 \hat{b}_i^{\rm (in)}|0,{\rm in}\rangle=0\,,\;\;\;\;\;\; \hat{b}_i^{\rm (out)}|0,{\rm out}\rangle=0\,
\end{eqnarray}
for particles, which we will denote symbolically by ``$i>F$'', where $F$ stands for the ``Fermi level'',  and
\begin{eqnarray}
 \hat{d}_i^{\rm (in)}|0,{\rm in}\rangle=0\,,\;\;\;\;\;\; \hat{d}_i^{\rm (out)}|0,{\rm out}\rangle=0\,
\end{eqnarray}
for antiparticles (``$i<F$'').
These operators
obey the standard anticommutation relations.
The electron-positron field operator
$\hat{\Psi}(\bfr,t)$ in the Heisenberg picture is defined as
\begin{eqnarray} \label{psi_in}
  \hat{\Psi}(\bfr,t)=\sum_{i>F} \hat{b}_i^{\rm (in)} \psi_i^{(+)}(\bfr, t) + \sum_{i<F} \hat{d}_i^{{\rm (in)}\dagger} \psi_i^{(+)}(\bfr, t)\,, \\
   \hat{\Psi}(\bfr,t)=\sum_{i>F} \hat{b}_i^{\rm (out)} \psi_i^{(-)}(\bfr, t) + \sum_{i<F} \hat{d}_i^{{\rm (out)}\dagger} \psi_i^{(-)}(\bfr, t)\,.
\label{psi_out}
\end{eqnarray}
For  collisions of bare nuclei, the initial  $t=t_{\rm in}$ state
of the system is described by the vacuum vector $|0,{\rm in}\rangle$.
To find the number of the electrons in a state ``$k$'' at $t=t_{\rm out}$, one should evaluate the value
$n_k=\langle 0,{\rm in}|\hat{b}_k^{{\rm (out)}\dagger}\hat{b}_k^{\rm (out)}|0,{\rm in}\rangle$.
This can be done by employing the equations
(\ref{psi_in})-(\ref{psi_out}) and the anticommutation relations between the creation and annihilation operators.
As the result, one obtains \cite{gre85,fra91}
\begin{eqnarray}
n_k=\langle 0,{\rm in}|\hat{b}_k^{{\rm (out)}\dagger}\hat{b}_k^{\rm (out)}|0,{\rm in}\rangle = \sum_{i<F}|a_{ki}|^2\,,
\end{eqnarray}
where
\begin{eqnarray}
a_{ij}= \int d \bfr \psi_i^{(-)\dagger}(\bfr, t) \psi_j^{(+)}(\bfr, t)
\end{eqnarray}
is the one-electron  transition amplitude.
Since the amplitudes $a_{ij}$ are time-independent, they can be evaluated at
the time moment~$t_{\rm in}$ or  $t_{\rm out}$ :
\begin{eqnarray}
  a_{ij} &=& \int d \bfr \; \psi_i^{(-){\dag}}(\bm{r},t_{\rm in}) \phi_j^{\rm in}(\bm r)\nonumber\\
  &=& \int d \bfr \; \phi_i^{\rm out{\dag}}(\bm{r}) \psi_j^{(+)}(\bm r,t_{\rm out})\,.
\end{eqnarray}
Alternatively, one can calculate the number of created positrons in a state ``$p$'' by formula
\begin{eqnarray}\label{n_p}
  \overline{n}_p=\langle 0,{\rm in}|\hat{d}_p^{{\rm (out)}\dagger}\hat{d}_p^{\rm (out)}|0,{\rm in}\rangle
  = \sum_{i>F}|a_{pi}|^2\,.
\end{eqnarray}
The total number of the electron-positron pairs can be found as
\begin{eqnarray}\label{p_tot}
P=\sum_{k>F} n_k =\sum_{p<F}\overline{n}_p\,.
\end{eqnarray}
Taking into account that for the processes under consideration  $P \ll 1$, we will
refer to this value as the pair-creation probability.

For the calculations of $n_k$ and $\overline{n}_p$ it is convenient to use the finite basis set method
\cite{dey12,dey13,mal15,bon15,mal17,pop18,mal18}.
With this method, one gets a spectrum consisting of
a finite number of states,
including both  bound and continuum (positive- and negative-energy) pseudostates.
As a result,
all the related summations in the  equations given above run over a finite number of states.

Finally, it should be emphasized  that the formulas presented in this section account
for all pair-creation channels ($a$, $b$, $c$, and $d$) shown in
Fig.~\ref{fig:energy_mol}.

\subsection{Monopole approximation}

Numerical solution of the TDDE with the total two-center potential (\ref{pot2}) is very
time consuming. For this reason, the calculations of the pair-creation probabilities are
generally restricted to the so-called monopole approximation for $V(\bfr,t)$.  In this approximation,
the potential   $V(\bfr,t)$  is expanded in spherical harmonics in the center-of-mass frame
and only the lowest-order spherical-symmetric term of this expansion is taken into account,
\begin{eqnarray} \label{pot_mon}
  V^{\rm (mon)}(r,t) =\frac{1}{4\pi}\int d{\bfn}\; V(\bfr,t)\,,
\end{eqnarray}
where ${\bfn}=\bfr/r$.
The direct calculations of the pair-creation probabilities
 and the positron spectra near their maxima
for the two-center potential
\cite{mal17,pop18,mal18} have demonstrated that the monopole approximation works rather well,
unless the impact parameter of the nuclear collisions is too large. In particular,
in the case of uranium-uranium collisions at the projectile energy $6.2$ MeV/u
(in the rest frame of the target nucleus before the collision)
the difference between the two-center and
the monopole-approximation
results  for the total probability  varies from about 6\%
for the head-on collision ($b = 0$)
to about 10\% at $b=10$ fm. Since
the impact parameter  $b=10$ fm corresponds to the scattering angle $\theta= 79^o$
in the center-of-mass frame
(at the energy  under consideration), this
means the validity of the monopole approximation
is preserved
in a wide range of the scattering
angles around the backward direction.

For the spherically symmetric potential (\ref{pot_mon}),  the Dirac wave functions
are represented in the standard form
\begin{eqnarray}
{\psi_{\kappa m}(\bfr,t)} =
\frac{1}{r}\left( \begin{array}{c}
 {G_{\kappa}(r,t)}\Omega_{\kappa m}(\bfn)\\
i{F_{\kappa}(r,t)}\Omega_{-\kappa m}(\bfn)
\end{array} \right)
\,,
\end{eqnarray}
where $ \Omega_{\kappa m}(\bfn) $ is the spherical spinor,
$G_{\kappa}(r,t)/{r}$ and $F_{\kappa}(r,t)/{r}$ are the radial Dirac components,
    and  $\kappa=(-1)^{j +l+1/2}(j+1/2)$ is the relativistic quantum number defined by
    the angular momentum and parity.  Due to the conservation of the total angular momentum,
    one has to deal with the time-dependent radial Dirac equation for a given value
    of $\kappa$,
 \begin{eqnarray} \label{dir_rad}
   i\frac{\partial}{\partial t} \phi(r,t) = H_{\kappa}(t)\phi(r,t),
 \end{eqnarray}
 where
 \begin{eqnarray}
{\phi(r,t)} =
\left( \begin{array}{c}
 {G(r,t)}\\
 {F(r,t)}
\end{array} \right)
\,
\end{eqnarray}
is the radial Dirac wave function and
 \begin{eqnarray}
{H_{\kappa }(t)} =
\left( \begin{array}{cc}
 m_e+V^{\rm (mon)}(r,t) & -\frac{d}{dr}+\frac{\kappa}{r} \\
  \frac{d}{dr}+\frac{\kappa}{r}  & -m_e+ V^{\rm (mon)}(r,t)
\end{array} \right)
\,
\end{eqnarray}
 is the radial Dirac Hamiltonian.

 For a given $\kappa$,  the initial states,
 including the bound and  continuum pseudostates, are
 obtained by diagonalization of the matrix
 ${\cal H} \equiv H_{\kappa}(t_{\rm in})$
 in a finite basis set.
The basis functions are constructed from B-splines \cite{sap96}
within the framework of the dual-kinetic-balance approach \cite{sha04}.
To solve Eq. (\ref{dir_rad}), we expand $\phi(r,t)$
on a basis of the eigenstates of the matrix $\mathcal{H}$.
For a given initial condition defined by Eq. (\ref{init_cond}),
we have
\begin{align}
\label{eqn:PhiDecomp}
  \phi_i({r},t) = \sum_{k = 1}^N c_{ki}(t)u_k({r})e^{-i\varepsilon_kt},
\end{align}
where $N$ is the number of the states,
$\varepsilon_k$ are the eigenvalues of the ${\cal H}$ matrix,
and $c_{ki}$ are the expansion coefficients. Substitution of the expansion \eqref{eqn:PhiDecomp}
into Eq. (\ref{dir_rad})
leads to the equations:
\begin{align}
  \label{eqn:SDEq}
    i \frac{\partial}{\partial t} c_{ji}(t) = \sum_k V_{jk}(t)c_{ki}(t), \text{\ \ \ subject to } c_{ji}(t_{\rm in}) = \delta_{ji},
\end{align}
where
\begin{align}
  \label{eqn:Vij}
  V_{jk}(t) = \langle u_j |(V^{\rm (mon)}(r,t) - V^{\rm (mon)}(r,t_{\rm in}))| u_k \rangle  e^{-i(\varepsilon_k - \varepsilon_j)t}.
\end{align}
The equations~\eqref{eqn:SDEq} are solved using the Crank-Nicolson scheme~\cite{cra47}:
\begin{align}
  {\vec c}_i(t + \Delta t) \approx M(t + \Delta t; t){\vec c}_{i}(t),
\end{align}
where $\Delta t$ is a small time step,
$\vec c_i = \{c_{1i}, \dots, c_{Ni} \}$,
and the $M$ matrix is determined by
\begin{eqnarray}
  M(t + \Delta t; t)& =& \left[I + i \frac{\Delta t}{2}V(t + \frac{\Delta t}{2})\right]^{-1} \nonumber \\
  && \times  \left[I - i \frac{\Delta t}{2}V(t + \frac{\Delta t}{2})\right].
\end{eqnarray}
With this technique we propagate all the bound and continuum quasistates
from $t_{\rm in}$ to $t_{\rm out}$ and calculate the pair-creation probability
for a given $\kappa$ using the formula (\ref{p_tot}). The total pair-creation probability
is obtained as a sum of the partial $\kappa$ contributions.
Usually it is sufficient to
account for the $ \kappa =\pm 1 $ contributions only. This is due to
the dominant contribution of the pair-creation processes in which
the created electrons are captured into bound states \cite{mul88,mal15,pop18}.

The calculation of the positron spectrum can be performed according to Eq. (\ref{n_p}).
However, due to the use of the finite basis set methods, the direct calculation
of the energy-differential spectrum $dP/d\varepsilon$ by this
formula is not possible.
Therefore, to obtain $dP/d\varepsilon_p$ we use
the Stieltjes method \cite{lan74,mal15,bon15}:
\begin{eqnarray}
  \frac{dP}{d\varepsilon}\Bigl(\frac{\varepsilon_p+\varepsilon_{p+1}}{2}\Bigr)
  =\frac{1}{2}\frac{\overline{n}_{p+1}+\overline{n}_{p}}{\varepsilon_{p+1}-\varepsilon_{p}}\,,
\end{eqnarray}
where $\varepsilon_{p}$ is the eigenvalue of the Hamiltonian in the finite basis set.

\subsection {Choice of trajectories}

\begin{figure}
\begin{center}
  \includegraphics[width=\columnwidth]{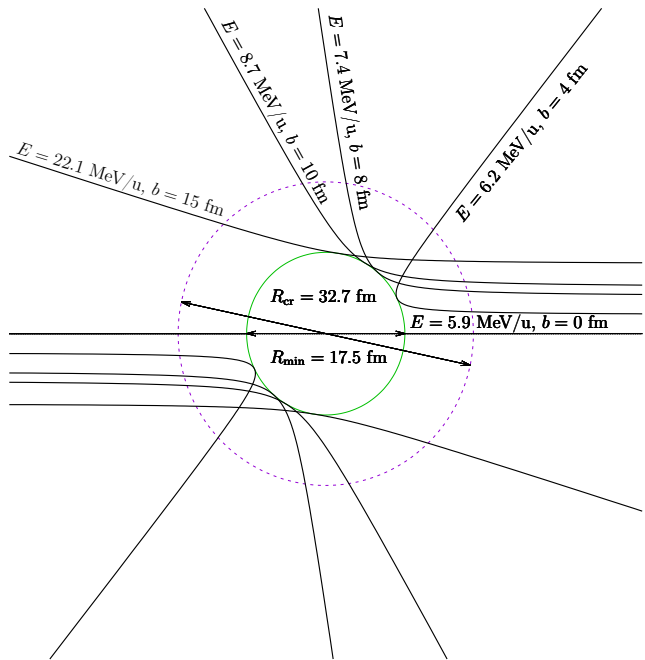}
\end{center}
\caption{The  U$^{92+}$ $-$ U$^{92+}$ collision trajectories (in the center-of-mass frame)
which correspond to the same minimal distance  $R_{\rm min}=17.5$ fm for different
impact parameters $b$ and, therefore, for different collision energies $E$.
The energies $E$ denote the projectile kinetic energies in
the rest frame of the target nucleus before the collision.
The circle  $R_{\rm min}=32.7$ fm defines the boundary (critical distance)
between the subcritical and the supercritical regimes.}
\label{fig:tr}
\end{figure}

\begin{figure}
\begin{center}
  \includegraphics[width=\columnwidth]{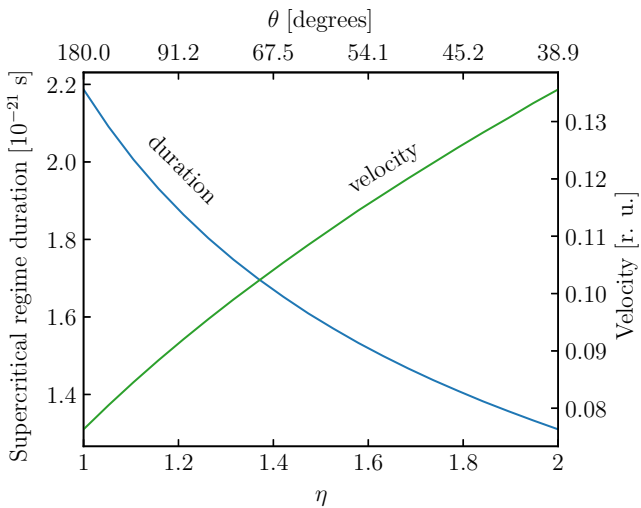}
\end{center}
\caption{The supercritical regime duration and the relative velocity of the nuclei
  at the $R= R_{\rm cr}$  in U$^{92+}$ $-$ U$^{92+}$  collisions at  $R_{\rm min}=17.5$ fm (see Fig.~\ref{fig:tr})
  as functions of the collision energy, $\eta=E/E_0$,
  where $E_0$ is the head-on collision energy at the same $R_{\rm min}$. The scattering
  angles, $\theta$, which correspond to the given values of $\eta$, are also indicated.}
\label{fig:time}
\end{figure}

In what follows, we will mainly consider the collisions of bare nuclei
at the energies close to the Coulomb barrier, which is defined as the
collision energy at which the nuclei touch each other.
The nuclear trajectories are defined by nonrelativisic classical mechanics
and, in the case of bare nuclei, are given by
the well-known
equations \cite{landau1}.
According to these equations, the minimal distance between the nuclei, $R_{\rm min}$,
is related to the impact parameter $b$ by
\begin{eqnarray}
  b^2=R_{\rm min}^2 -\frac{\alpha Z_1Z_2} {E} R_{\rm min}\,,
 \end{eqnarray}
where $E$ is the collision energy in the center-of-mass frame.
For a given value of $ R_{\rm min}$,
the minimal energy, $E_0$,  corresponds to the head-on ($b=0$) collision,
\begin{eqnarray}
E_0=\frac{\alpha Z_1Z_2}{R_{\rm min}}.
\end{eqnarray}
Let us  consider only
the trajectories which correspond to the same minimal distance ($R_{\rm min}$)
for different impact parameters ($b$) and, therefore, for different collision energies ($E$).
In case of uranium-uranium collisions with  $R_{\rm min}=17.5$ fm,
which corresponds to the projectile energy $E_0=5.9$ MeV/u (in the rest frame
of the target nucleus before the collision),
the trajectories of interest
are shown in  Fig.~\ref{fig:tr}
in the center-of-mass frame of reference.
In addition to the circle, corresponding to  $R_{\rm min}=17.5$ fm,
we display also the circle with  $R_{\rm cr}=32.7$ fm, which determines the boundary
(critical distance) between
the subcritical and the supercritical regimes. Thus, we have the supercritical regime
only during the period of time when the nuclei move from $R_{\rm cr}$ to
$R_{\rm min}$ and back. This period of time decreases with increasing
the collision energy ($E$) and, therefore, with increasing
the impact parameter $b$ (within the
trajectories under consideration). This fact is clearly demonstrated in Fig.~\ref{fig:time}, which
shows the supercritical regime duration as a function of $\eta=E/E_0$ at fixed
$R_{\rm min}=17.5$ fm (for all $b$ and $E$).
For convenience, the figure also shows the scattering angles, $\theta$, which
correspond to the given values of $\eta$.
In the same figure, we present also the relative velocity of the nuclei
at the distance $R= R_{\rm cr}$ as a function of $E/E_0$.
Opposite to the  supercritical regime duration,
this velocity increases with increasing  $E/E_0$.
It is clear that  the dynamical pair creation must decrease monotonously  with decreasing
velocity (and, therefore, $E/E_0$) for a given $R_{\rm min}$. As to the spontaneous pair creation,
it must increase monotonously
with increasing the supercritical regime duration and, therefore,
with decreasing $E/E_0$. It follows that any increase in the pair-production
probability at  $E/E_0 \rightarrow 1$ for a given $R_{\rm min}$
should indicate the effect of the spontaneous pair creation,
which takes place in the supercritical regime only.

\section{Calculations and results}

The methods described in the previous section are employed for the calculations of the total
pair-creation probabilities and positron spectra. The calculations are performed within the framework
of the monopole approximation and are restricted to the $|\kappa|=1$ contributions.
As is noted above, these contributions almost completely
determine the values of interest and
the difference between the monopole approximation
results and the full two-center results does not exceed 10\% in a wide interval of the scattering angles.

\begin{figure}
\begin{center}
  \includegraphics[width=\columnwidth]{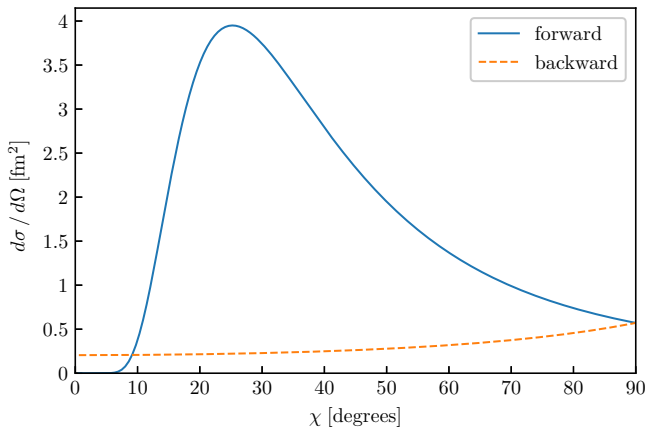}
\end{center}
\caption{The pair-creation cross sections as functions
  of $\chi=180^o-\theta$ for the backward ($\theta > 90^o$)
  scattering trajectories and $\chi=\theta$ for
  the forward  ($\theta < 90^o$)
  scattering trajectories.}
\label{fig:3}
\end{figure}
\begin{figure}
  \begin{center}
  \includegraphics[width=\columnwidth]{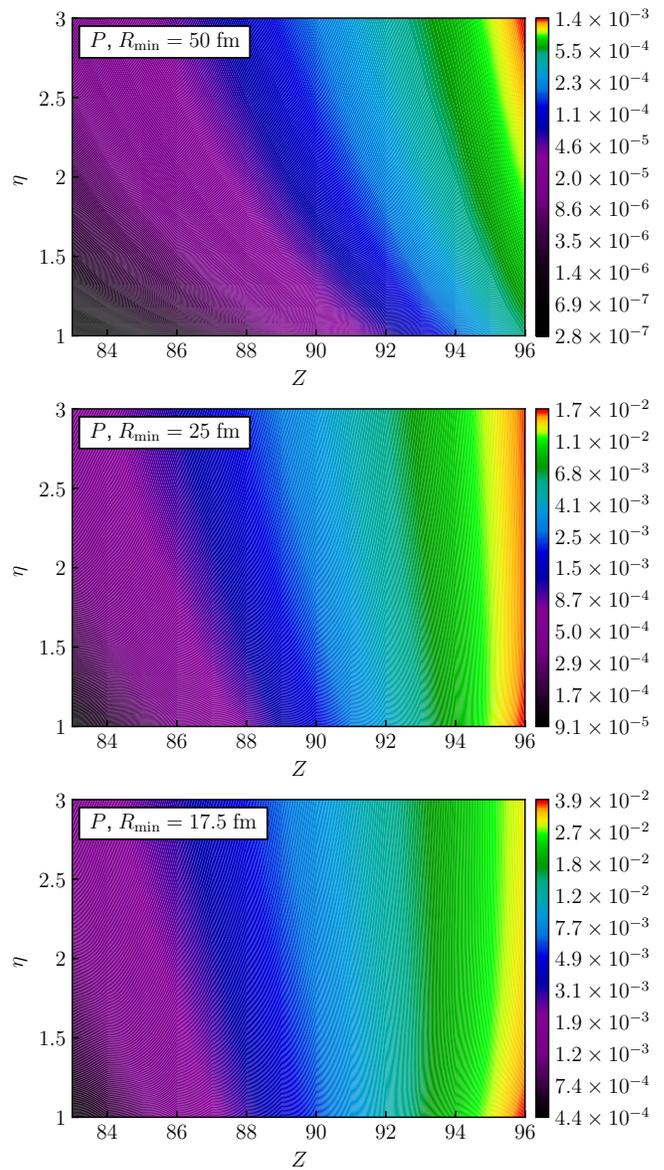}
\end{center}
  \caption{The pair-creation probability in symmetric ($Z=Z_1=Z_2$) collisions
    as a function
of the parameter $\eta=E/E_0$, which is the ratio of the collision energy $E$
to the energy of the head-on collision $E_0$,
and the nuclear charge number $Z$
 at different minimal internuclear distance
 $R_{\rm min}$, which is the same within each image.
}
\label{fig:P_eta}
\end{figure}

\begin{figure}
  \begin{center}
    \includegraphics[width=\columnwidth]{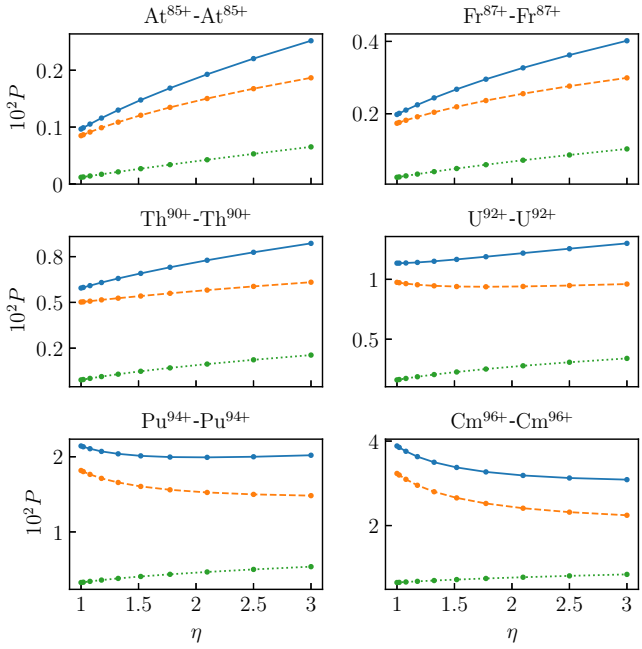}
\end{center}
  \caption{The  $s(\kappa=-1)$ and $p_{1/2}(\kappa=1)$ contributions to
  the  pair-creation probability $P$  as functions
of the parameter $\eta=E/E_0$, which is the ratio of the collision energy $E$
to the energy of the head-on collision $E_0$,
 at the same  minimal internuclear distance
$R_{\rm min}=17.5 $ fm. The $s$ and $p_{1/2}$ contributions are shown with
the dashed (orange)  and  dotted (green) lines, respectively, while the total
$P$ value is displayed by the solid (blue) line.
}
\label{fig:P_eta1}
\end{figure}
\begin{figure}
\begin{center}
  \includegraphics[width=\columnwidth]{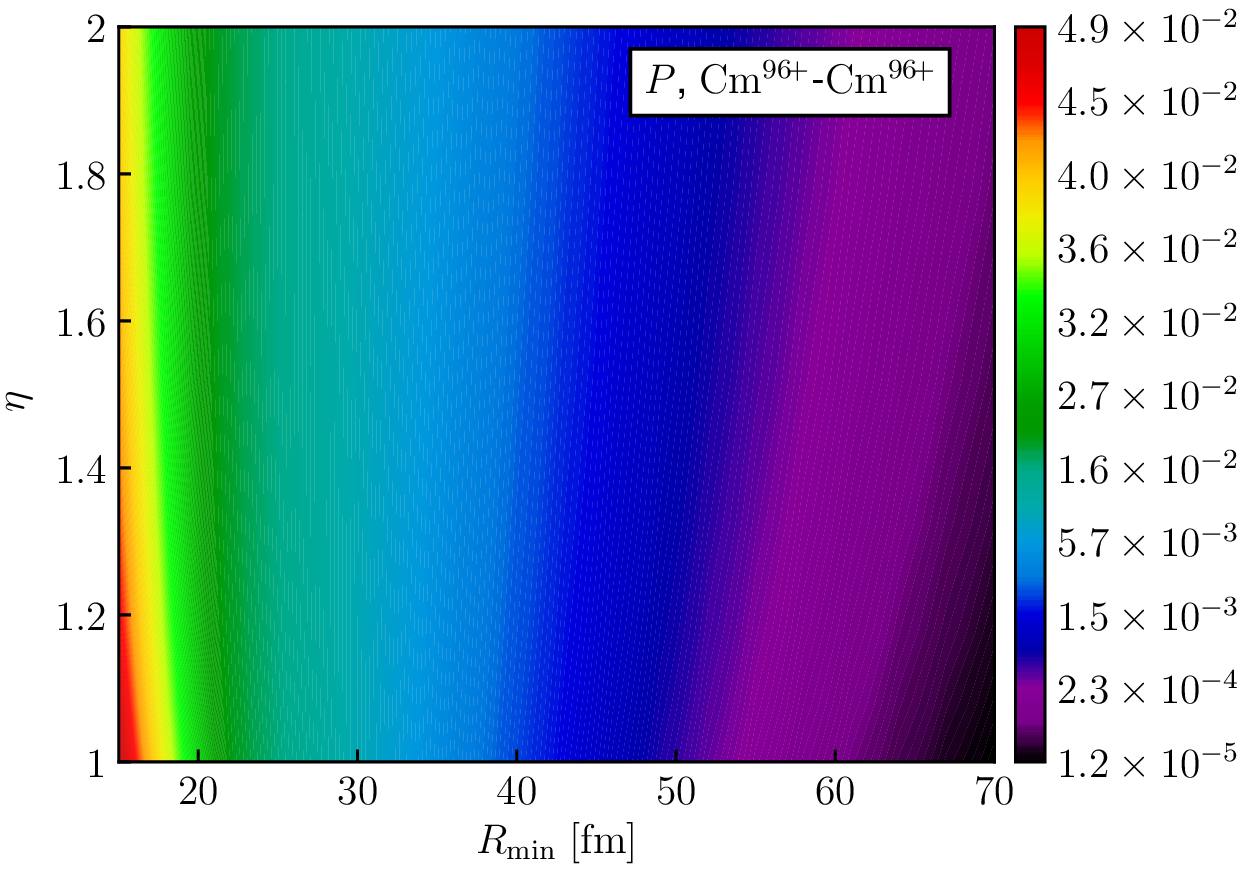}
\end{center}
\caption{The pair-creation probability in symmetric  $Z_1=Z_2=96$ collisions
    as a function of the minimal internuclear distance $R_{\rm min}$ and
the parameter $\eta=E/E_0$, which is the ratio of the collision energy $E$
to the energy of the head-on collision at the same $R_{\rm min}$.
}
\label{fig:P_rmin_96}
\end{figure}

\begin{figure}
\begin{center}
  \includegraphics[width=\columnwidth]{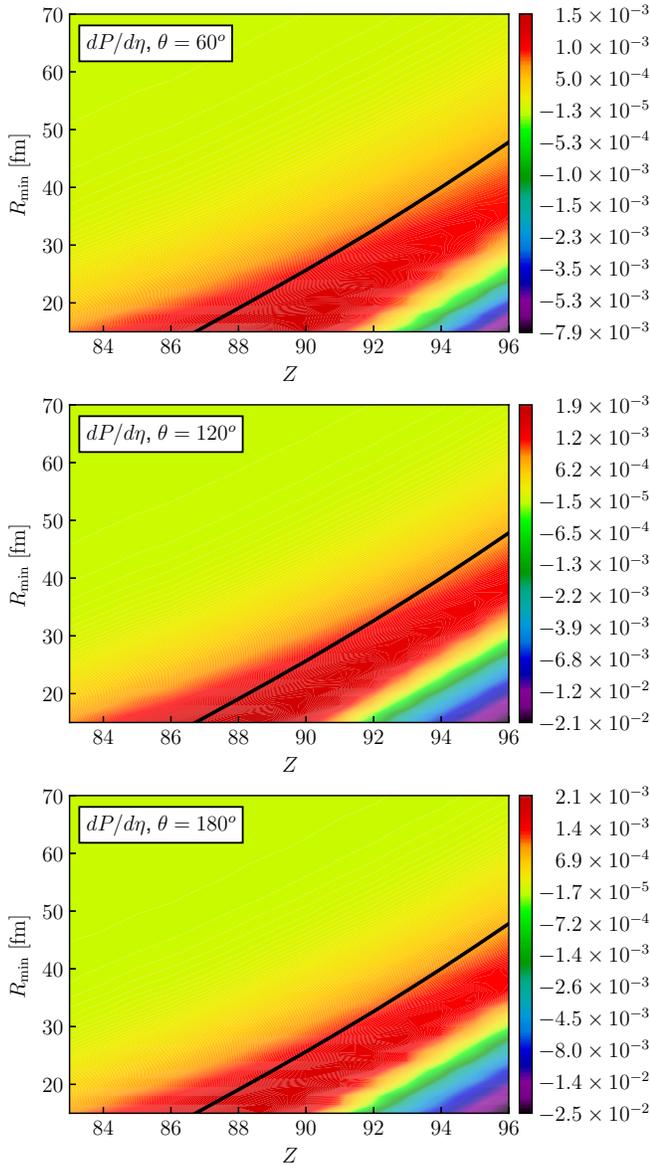}
\end{center}
\caption{The derivative of the pair-creation probability with respect to $\eta$,  $dP/d\eta$,
as a function
of $Z=Z_1=Z_2$ and $R_{\rm min}$.
The black line  indicates the boundary between the subcritical and the supercritical regimes.
}
\label{fig:dPdeta}
\end{figure}

\begin{figure}
\begin{center}
  \includegraphics[width=\columnwidth]{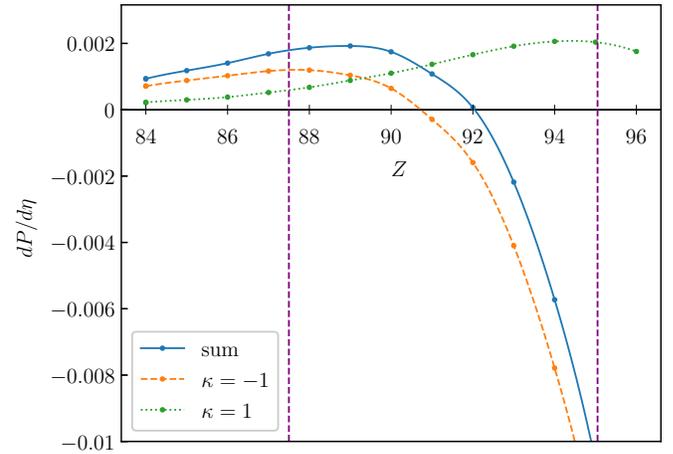}
\end{center}
\caption{The derivative of the pair-creation probability  $dP/d\eta$ at  $\eta=1$ ($\theta=180^o$)
as a function
of $Z=Z_1=Z_2$ for $R_{\rm min}=17.5$~fm.
The dashed vertical lines at $Z\approx 87.5$ and $Z\approx 95$
indicate the boundaries between the subcritical and the supercritical regimes
for the $1s$ ($\kappa=-1$) and $2p_{1/2}$ ($\kappa=1$)
states, respectively.}
\label{fig:der_pm}
\end{figure}

\begin{figure}
\begin{center}
  \includegraphics[width=\columnwidth]{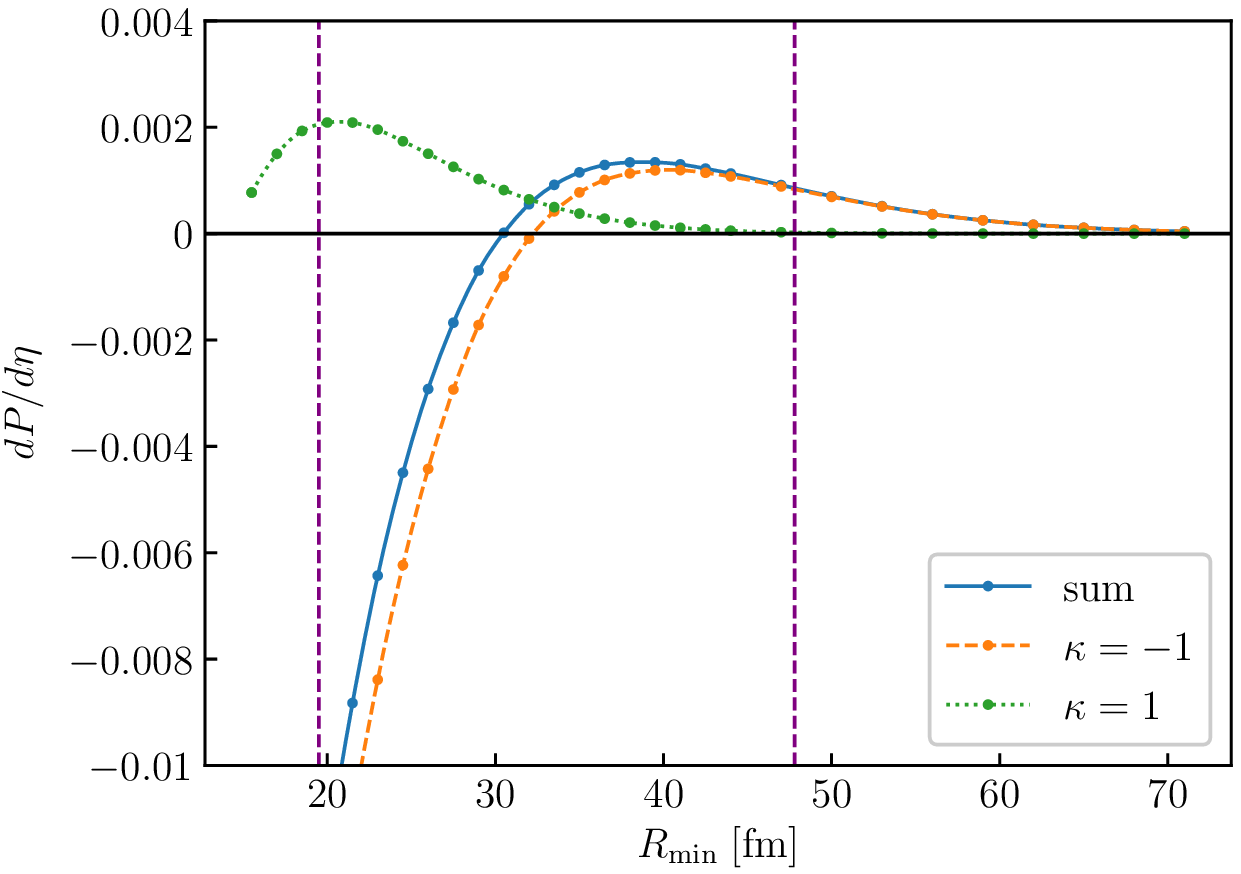}
\end{center}
\caption{The derivative of the pair-creation probability $dP/d\eta$ at $\eta=1$ ($\theta=180^o$)
as a function
of $R_{\rm min}$ for $Z_1=Z_2=96$.
The dashed vertical lines at
$R_{\rm min} = R_{\rm cr}^{1s}\approx 48$~fm and $R_{\rm min}=R_{\rm cr}^{2p_{1/2}}\approx 20$~fm
indicate the boundaries between the subcritical and the supercritical regimes
for the $1s$  ($\kappa=-1$) and $2p_{1/2}$  ($\kappa=1$) states, respectively.
}
\label{fig:dPdeta_rmin}
\end{figure}

The positron-creation probabilities and positron spectra in heavy-ion collisions can be
measured by requiring a coincidence between a positron event and
scattered projectiles or recoil target nuclei at given angles
\cite{bac78,koz79}.
If we neglect the nuclear-polarization effects and the oblate shape
of the nuclei under consideration (e.g., uranium),
there is a simple correspondence between the scattering
angle and the impact parameter. So, we can study the pair-creation
probabilities as functions of the impact parameter.
However, in case of fully symmetric collisions,
one should keep in mind a problem of distinguishing the contributions for the angles $\theta$ and $180^o -\theta$,
which correspond to different impact parameters. The calculations for the uranium-uranium collisions
at the energies near the Coulomb barrier show that the pair creation from
the backward scattering trajectories dominate over the pair creation from
the related forward scattering trajectories in a rather small region: $\theta = 180^o-172^o$. This
can be seen from  Fig.~\ref{fig:3} which compares the corresponding pair-creation cross sections as functions
of $\chi=180^o-\theta$ for the backward ($\theta > 90^o$)
scattering trajectories and $\chi=\theta$ for
the forward ($\theta < 90^o$) scattering trajectories.
This problem does not appear for non-symmetric collisions
as well as for collisions of bare nuclei with
neutral atoms, which will be briefly discussed at the end of
this section. However, in what follows, for simplicity we will
mainly consider the symmetric ($Z=Z_1=Z_2$) collisions
of bare nuclei assuming that, if necessary, the backward and forward
trajectories can be distinguished by colliding
different isotope beams. Some results for
non-symmetric collisions will be presented in Sect. IIID.

\begin{figure*}
\begin{center}
  \includegraphics[width=1.0\textwidth]{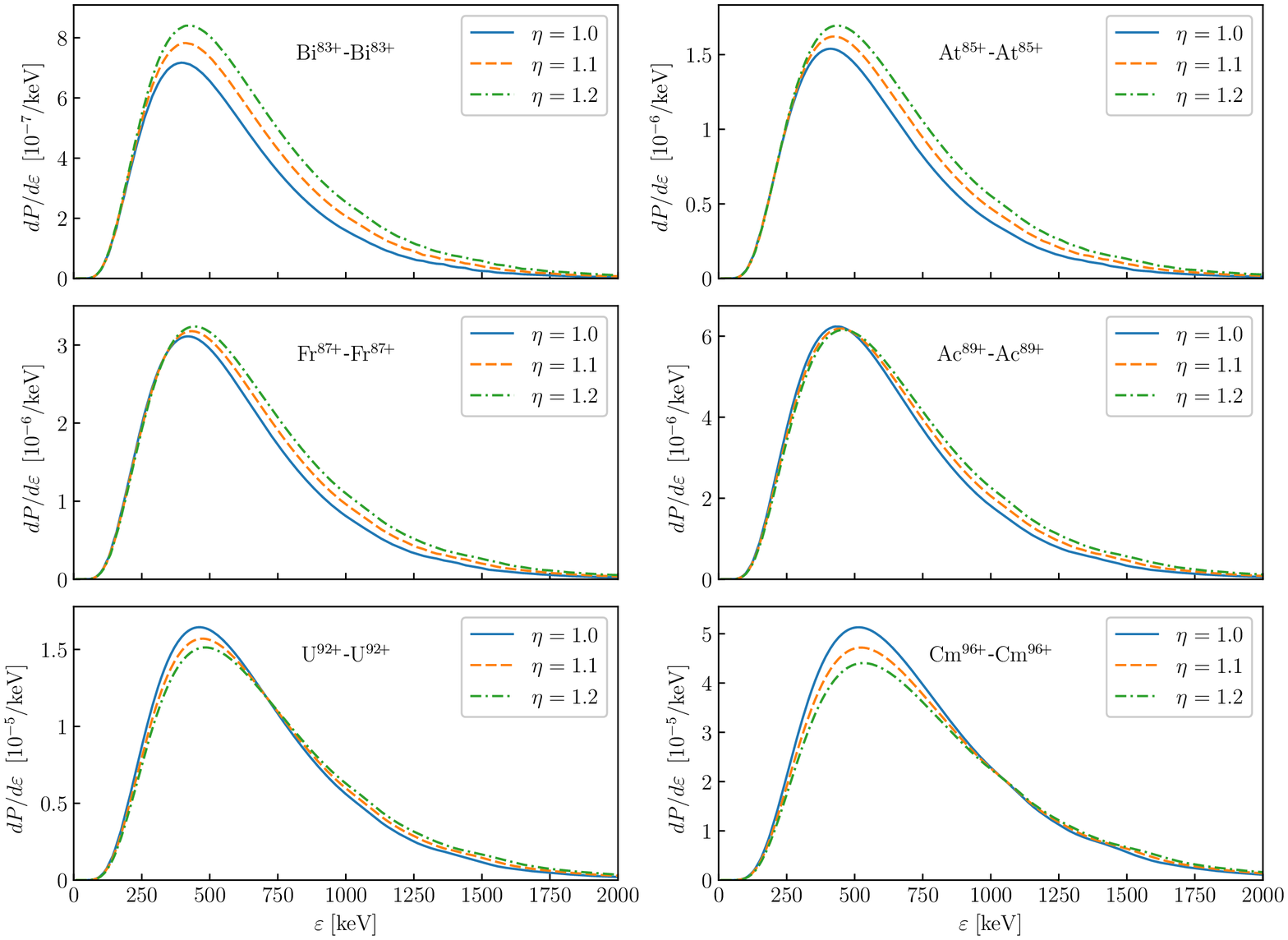}
\end{center}
\caption{The positron spectra for the symmetric collisions with $Z=Z_1=Z_2=83-96$
  at $R_{\rm min}=17.5$ fm and  $\eta=E/E_0=1.0, 1.1, 1.2$,
  which is the ratio of the collision energy $E$
to the energy of the head-on collision at the same $R_{\rm min}$.}
\label{fig:sp1}
\end{figure*}

\begin{figure*}
\begin{center}
  \includegraphics[width=1.0\textwidth]{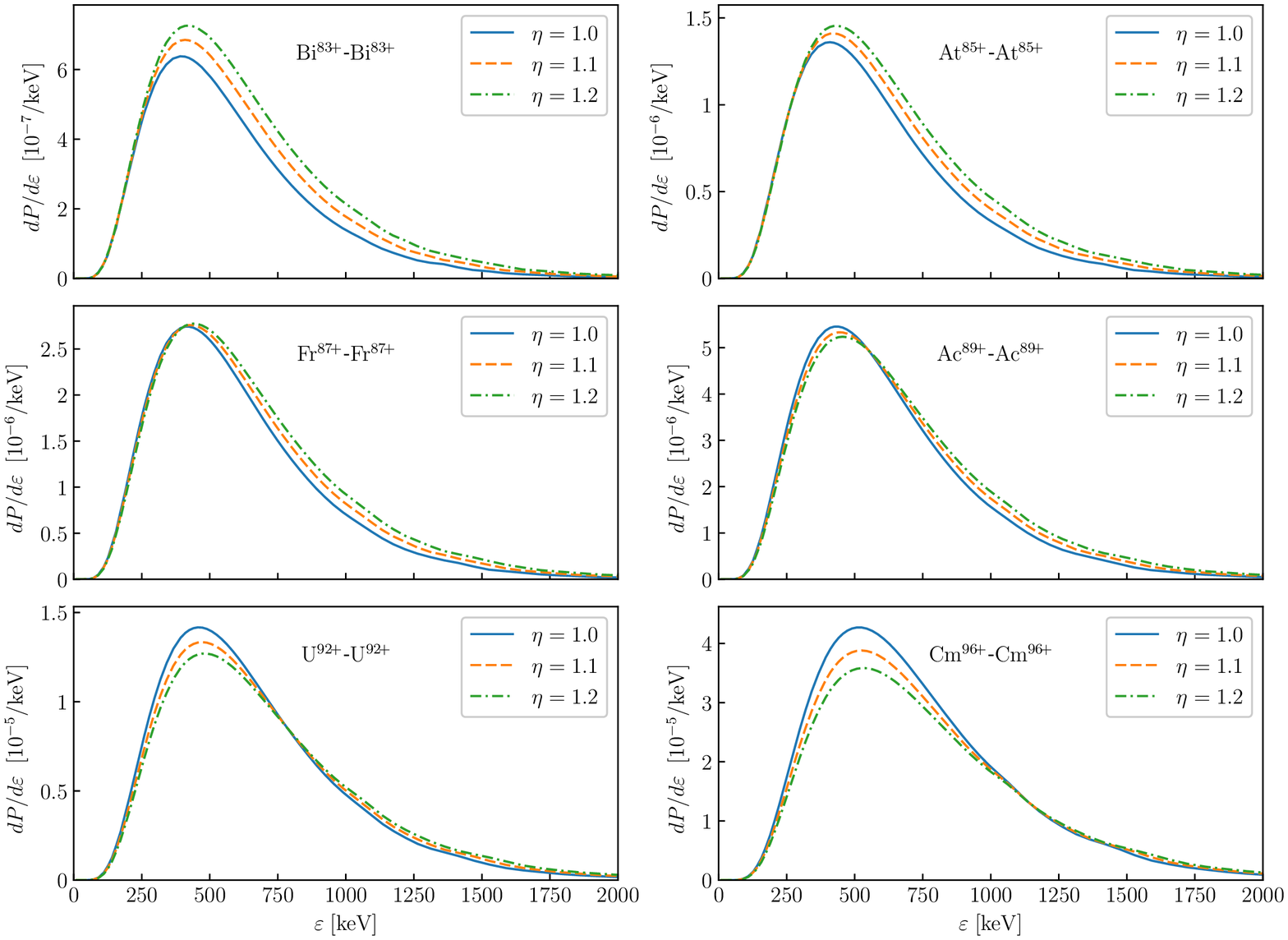}
\end{center}
\caption{The $s$-wave ($\kappa=-1$) contribution to the  positron spectra for the symmetric
  collisions with $Z=Z_1=Z_2=83-96$
  at $R_{\rm min}=17.5$ fm and  $\eta=E/E_0=1.0, 1.1, 1.2$,
   which is the ratio of the collision energy $E$
to the energy of the head-on collision at the same $R_{\rm min}$.
}
\label{fig:sp2}
\end{figure*}

\begin{figure*}
\begin{center}
  \includegraphics[width=1.0\textwidth]{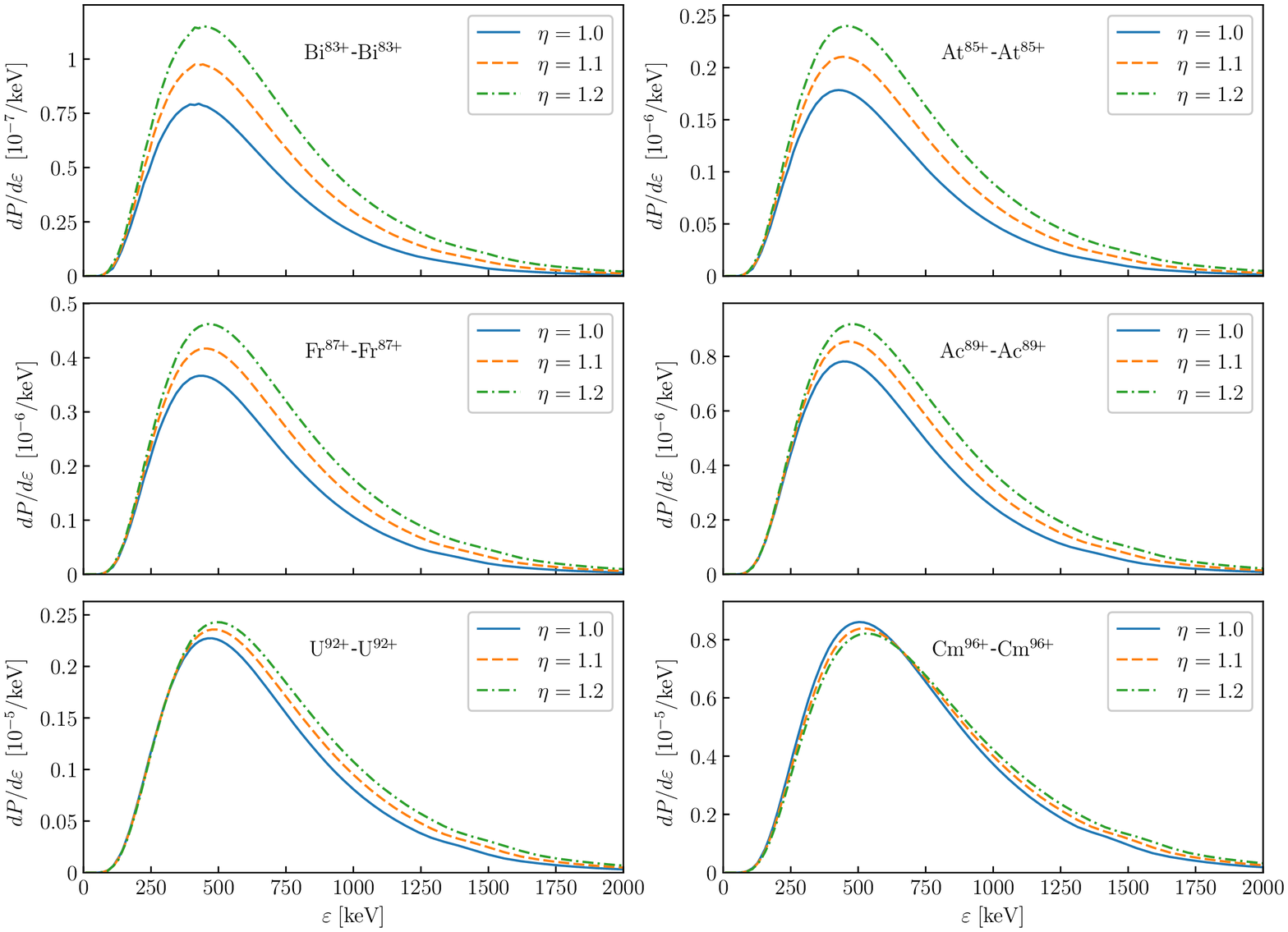}
\end{center}
\caption{The $p_{1/2}$-wave  ($\kappa=1$) contribution to the  positron spectra
  for the symmetric collisions with $Z=Z_1=Z_2=83-96$
  at $R_{\rm min}=17.5$ fm and  $\eta=E/E_0=1.0, 1.1, 1.2$,
   which is the ratio of the collision energy $E$
to the energy of the head-on collision at the same $R_{\rm min}$.
}
\label{fig:sp3}
\end{figure*}

\begin{figure*}
\begin{center}
  \includegraphics[width=1.0\textwidth] {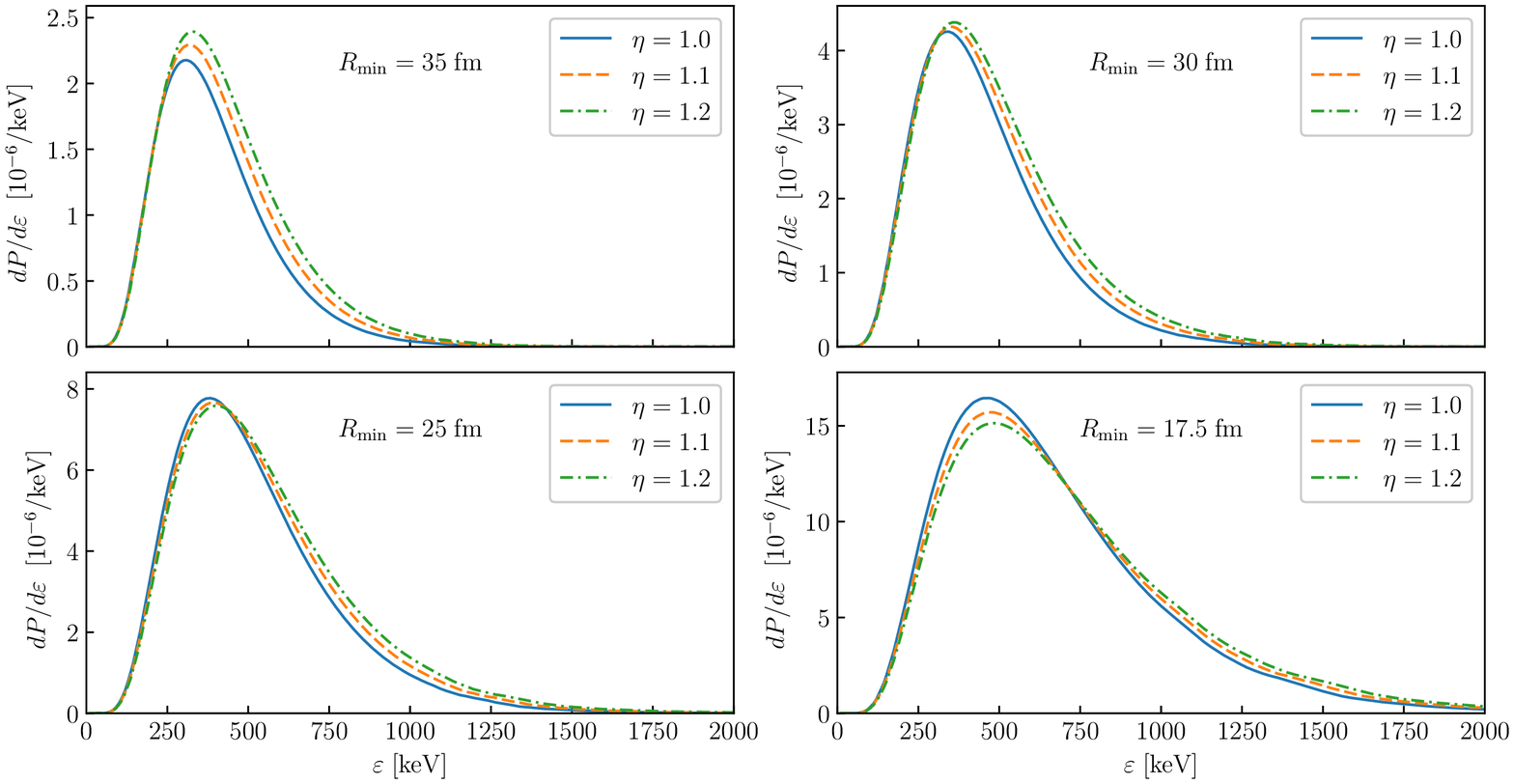}
\end{center}
\caption{The positron spectra in the U$^{92+}-$U$^{92+}$ collision
at different $R_{\rm min}$ and  $\eta=E/E_0$,
   which is the ratio of the collision energy $E$
to the energy of the head-on collision at the same $R_{\rm min}$.
}
\label{fig:spectrum92}
\end{figure*}

\subsection{Pair-creation probabilities for symmetric collisions}

In   Fig.~\ref{fig:P_eta} we present the results for the pair-creation probability in symmetric
($Z=Z_1=Z_2$) collisions as a function
of the parameter $\eta=E/E_0$, which is the ratio of the collision energy $E$
to the energy of the head-on collision $E_0$,
and the nuclear charge number $Z$
at the same minimal internuclear distance
$R_{\rm min}$ (it means that only the trajectories as presented in Fig.~\ref{fig:tr}  must
be considered for each $R_{\rm min}$).
The results are given
for the values of $R_{\rm min} =17.5, 25, 50$ fm, while
the critical distance for the largest $Z$ presented ($Z=96$) is about $48$ fm.
It means that in the case $R_{\rm min}=50$ fm for all $Z$ under consideration
there is no contribution from  the spontaneous
pair creation,  and, therefore, the pair creation is completely due to the dynamical (induced) origin.
As a result, with $R_{\rm min}=50$ fm
for all $Z=84-96$ the pair-creation probabilities
decrease monotonously
when $\eta\rightarrow 1$.
This is rather evident, since in the limit of
 adiabatically slow collisions the dynamical pair creation must be zero.
 The  behavior of $P$ changes when the supercritical regime becomes available.
 For $R_{\rm min}=17.5$ fm, in the subcritical regime
 ($Z<Z_{\rm cr}/2\approx 87.5$),
the pair-creation probability $P$ remains decreasing monotonously
when $\eta\rightarrow 1$. However,
the behavior of $P$ changes
when the nuclear charge number $Z$ becomes sufficiently higher than the critical value.
It can be seen that for such $Z$ the pair-creation probability $P$
increases at $\eta\rightarrow 1$.
This is due to the spontaneous pair-creation mechanism which starts to work
 in the supercritical regime,
while  the dynamical pair creation
decreases at $\eta\rightarrow 1$.
To demonstrate this behavior in more detail, in Fig.~\ref{fig:P_eta1}
we display the $s(\kappa=-1)$ and $p_{1/2}(\kappa=1)$ contributions to
the probability $P$ as functions of $\eta$ for
$Z=85-96$ at $R_{\rm min}=17.5$ fm (similar graphs for  $R_{\rm min}=16.5$ fm
have been presented in Ref. \cite{mal19}).
As one can see from this figure, the $s$ contribution
starts to increase significantly at  $\eta\rightarrow 1$ already for $Z=92$,
while the $p_{1/2}$ contribution
remains decreasing  at  $\eta\rightarrow 1$  even at $Z=96$. In the case of $Z=92$,
the increase of the $s$ contribution at  $\eta\rightarrow 1$
is fully compensated by the decrease of the $p_{1/2}$ contribution, that leads to
 an almost constant value of their sum at small $\eta$.
The different $\eta$ dependencies of the $s$ and $p_{1/2}$ contributions are explained
by different values of $Z$ at which the $1s$ and $2p_{1/2}$ states enter the negative-energy continuum.
For $R_{\rm min}=17.5$ fm, this takes place at  $Z=Z_{\rm cr}^{1s}/2\equiv Z_{\rm cr}/2\approx 87.5$ and
$Z=Z_{\rm cr}^{2p_{1/2}}/2\approx 95$ for  $1s$ and $2p_{1/2}$ states, respectively
(cf. Fig.~\ref{fig:energy_z}).

Fig.~\ref{fig:P_rmin_96}
displays the pair creation probabilities in symmetric $Z_1=Z_2=96$ collisions
as a function of  $R_{\rm min}$
and $\eta=E/E_0$, where $E_0$
corresponds to $R_{\rm min}$ under consideration.
As one can see, the behavior of $P$ at
$\eta\rightarrow 1$ changes when the minimal distance  $R_{\rm min}$ becomes
sufficiently smaller
than the critical value,  $R_{\rm cr}\approx 48$ fm for $Z=96$. In
the subcritical region, $R_{\rm min} > R_{\rm cr}$, the pair creation  is of
pure dynamical origin and, therefore, can only decrease with decreasing $\eta$
(at a given value of  $R_{\rm min}$). However, in the supercritical regime,
$R_{\rm min} < R_{\rm cr}$, the spontaneous pair creation becomes possible and
finally leads to an increase of $P$ when  $\eta \rightarrow 1 $.

Fig.~\ref{fig:dPdeta}
presents
the derivative of the pair-creation probability $P$ with respect to $\eta$,  $dP/d\eta$,
as a function
of $Z$ and $R_{\rm min}$ at different scattering angles (in the center-of-mass frame):
$\theta = 60^o,120^o,180^o$.
The black lines in this figure indicate the boundaries between the subcritical and the supercritical regimes.
According to our discussion above, the pictures in Fig.~\ref{fig:dPdeta}
clearly demonstrate that the $dP/d\eta$ derivative,
being strictly positive in the subcritical region, becomes negative when the parameters
$Z$ and $R_{\rm min}$ enter deeply enough into the supercritical domain. Moreover,
Fig.~\ref{fig:dPdeta} shows that the transition from the  subcritical to the
supercritical region  changes the behavior  of  $dP/d\eta$, starting  with an increase when approaching
the black line from the subcritical region and ending with a strong decrease after crossing this line.
It can be seen  that in
the strong supercritical regime
the $dP/d\eta$ derivative decreases in the directions which
correspond to increasing $Z$ and decreasing $R_{\rm min}$.
The profile   $R_{\rm min}=16.5$ fm, which
demonstrates the change of the behavior of $dP/d\eta$
from the original increase at $Z<Z_{\rm cr}/2$
to the subsequent  decrease at $Z>Z_{\rm cr}/2$,
has been presented for $\eta=1$ ($\theta=180^o$)
in Ref. \cite{mal19}.
In Fig.~\ref{fig:der_pm} we present separately the $\kappa=\pm 1$
contributions to  $dP/d\eta$ at  $\eta=1$ ($\theta=180^o$)
for the plane $R_{\rm min}=17.5$ fm.
It can be seen that,
while the $\kappa=-1$ contribution has a maximum at $Z\approx 88$,
the $\kappa=1$ contribution
has it at  $Z\approx 94.5$. This is due to a higher
value of the critical charge for the $2p_{1/2}$ state, $Z_{\rm cr}^{2p_{1/2}}/2 \approx 95$,
compared to the  $1s$ critical charge,
$Z_{\rm cr}^{1s}/2 \approx 87.5$, (cf. Fig.~\ref{fig:energy_z}).
As a result, the maximum of the total contribution
is shifted to $Z\approx 89$.
The profile $Z=96$ at  $\eta=1$ ($\theta=180^o$)
in Fig.~\ref{fig:dPdeta},
 which is separately
presented in  Fig.~\ref{fig:dPdeta_rmin}, shows a similar behavior
of $dP/d\eta$, when $R_{\rm min}$ passes the critical values,
$R_{\rm cr}^{1s}\approx 48$ fm and
$R_{\rm cr}^{2p_{1/2}}\approx 20$ fm for $Z=96$.

\begin{table}
  \caption{Supercritical resonance energies $\varepsilon_{\rm res}$  and
    related kinetic positron energies $\varepsilon=|\varepsilon_{\rm res}|-m_ec^2$
    for the ${\rm U}^{92+}-{\rm U}^{92+}$ and ${\rm Cm}^{96+}-{\rm Cm}^{96+}$
    systems at different internuclear distances $R$
    calculated
    within the framework of the monopole approximation.}
\label{isotope_shift}
\begin{center}
  \begin{tabular}{|c|lc|lc|lc|}\hline
    & \multicolumn{2}{c|} { ${\rm U}^{92+}-{\rm U}^{92+}$ }
    & \multicolumn{4}{c|} {${\rm Cm}^{96+}-{\rm Cm}^{96+}$} \\ \hline
     & \multicolumn{2}{c|} {$s$ resonance} & \multicolumn{2}{c|} {$s$ resonance} &  \multicolumn{2}{c|}
    {$p_{1/2}$ resonance}\\ \hline
    $R$ (fm)  & $\varepsilon_{\rm res}/m_ec^2$ & $\varepsilon $  (keV)& $\varepsilon_{\rm res}/m_ec^2$ & $\varepsilon $ (keV)
   & $\varepsilon_{\rm res}/m_ec^2$ & $\varepsilon $ (keV) \\
  \hline
  16   &  $-$1.582& 298 & $-$2.154 & 589 & $-$1.228  & 116\\
  18   &  $-$1.482& 246 & $-$2.011 & 517 & $-$1.092 & 47 \\
  20   &  $-$1.392& 200 & $-$1.885 & 452 & &\\
  22   &  $-$1.311& 159 & $-$1.773 & 395  & &\\
  24   &  $-$1.239& 122 & $-$1.673 & 344 & &\\
  26   &  $-$1.174& 89  & $-$1.584 & 298 & &\\
 \hline
\end{tabular}
\end{center}
\label{tab1}
\end{table}

\subsection{Positron spectra for symmetric collisions}

We have also calculated the positron energy distributions (in the
center-of-mass frame)
for the trajectories under consideration (Fig.~\ref{fig:tr}).
In Fig.~\ref{fig:sp1}
we present the total positron spectra for the symmetric collisions with $Z=Z_1=Z_2=83-96$
at $R_{\rm min}=17.5$ fm and  $\eta=E/E_0= 1.0, 1.1, 1.2$. In Figs.~\ref{fig:sp2} and~\ref{fig:sp3}
we display separately the contributions to the positron spectra from the $s (\kappa=-1)$ and
$p_{1/2}(\kappa=1)$ channels, respectively. As mentioned above, these two channels almost completely
determine the total probability. The presented positron spectra demonstrate the well known fact
(see, e.g., Refs. \cite{sof77,rei81}) that the emission of low-energy positrons
is strongly suppressed by the repulsive interaction with the nuclei
while at high energy
the spectra fall off exponentially.
In the case of $R_{\rm min}=17.5$ fm, the supercritical regime
starts at $Z\approx 87.5$ and $Z\approx 95$ for the $1s$ and $2p_{1/2}$ states, respectively.
As one can see from Figs.~\ref{fig:sp1},~\ref{fig:sp2}, and~\ref{fig:sp3},
in all cases the behavior
of the positron spectra manifests a clear signature of the transition to the supercritical
regime.
Indeed, in the subcritical regime
the positron spectrum curves at the energies near the maximum and higher
are strictly ordered according to  the $\eta$ values: the curve with a larger $\eta$ is strictly
higher than the curve with a smaller $\eta$. This is due to the fact that the dynamical
(induced) creation of positrons in general
should increase with increasing the
collision energy. In the supercritical regime, however,  in the region near the
maximum of the  positron spectrum,
the curves corresponding to smaller values of $\eta$ become higher than the ones
with larger $\eta$, while at sufficiently large  values of the positron energy  the order
of the curves remains the same as in the subcritical regime.
This might be explained as follows.
The spontaneous pair creation, which starts to work in
the supercritical regime, is greater at lower values
of $\eta$, which correspond to larger values of the  supercritical time duration
(Fig.~\ref{fig:time}).
In addition, the spontaneous positrons are restricted to
the lower part of the energy spectrum.
For instance, in the case of the U$^{92+}-$U$^{92+}$ collision,
 the spontaneous positrons
 should be mainly limited to the energy range 0 to 600~keV.
Indeed, let us consider first hypothetical collisions where
the nuclei move adiabatically slowly along
the actual trajectories \cite{mal15,mal19}.
In such collisions all created positrons are of pure spontaneous origin
and, therefore,
the energies of the emitted positrons are completely determined by the positions of the
supercritical resonances.
Table~\ref{tab1} presents these resonance energies, $\varepsilon_{\rm res}$,
and the related kinetic positron energies, $\varepsilon=|\varepsilon_{\rm res}|-m_ec^2$,
for the U$^{92+}-$U$^{92+}$ and Cm$^{96+}-$Cm$^{96+}$ systems at different internuclear distances.
These energies have been obtained within the framework of the monopole approximation employing
the complex rotation method \cite{moi98,lin12,ack07a,hor11,zay19}. The application of this
method for  the case of point nuclei leads to the results which are in good agreement
with the values obtained by the use of related formulas from Ref. \cite{god17}.
We note also that the corresponding calculations beyond the
monopole approximation \cite{hor11,mal20} yield the resonance energies
which are very close to the monopole-approximation ones. According to Table~\ref{tab1},
for the    U$^{92+}-$U$^{92+}$  system the positron kinetic energies do not exceed
300 keV, while the related natural resonance widths do not exceed a few keV \cite{hor11,mal20}.
But in the real (non-adiabatic) collision the finite  collision time yields
a dynamical width,
which is much larger than the natural resonance one. The dynamical width
can be roughly estimated by the uncertainty principle $\Gamma_{\rm dyn}\sim \hbar/\tau_{\rm col}$,
where $\tau_{\rm col}$ is determined as the supercritical time duration. With
$\tau_{\rm col}\sim 2\times10^{-21}$~s (see Fig.~\ref{fig:time}) one obtains
$\Gamma_{\rm dyn}\sim 300$ keV, which restricts the energy of spontaneous positrons
to the interval 0 to 600~keV.
The large value of  $\Gamma_{\rm dyn}$ prevents the existence of resonance structure
in the positron spectra and smears the area,
where the reverse order of the $\eta=1.0, 1.1, 1.2$
curves takes place, compared to the subcritical case.
In addition, the strong superposition of the dynamical and spontaneous
pair-creation mechanisms in the real collisions makes the transition
from the subcritical to the supercritical mode quite smooth.
However, the qualitative changes in the positron spectra,
which become especially convincing when comparing the
subcritical Bi$^{83+}-$Bi$^{83+}$ and supercritical Cm$^{96+}-$Cm$^{96+}$
collisions, must be considered as a conclusive
proof of the access to the supercritical mode.
This statement is also confirmed by  Fig.~\ref{fig:spectrum92}, where the positron
spectra for  the U$^{92+ }-$U$^{92+}$ collision at different $R_{\rm min}$ are presented.
Again, the transition from the subcritical to the supercritical mode ($R_{\rm cr}=32.7$ fm)
changes the order of the curves near the maximum.

\begin{figure}
\begin{center}
 \includegraphics[width=\columnwidth]{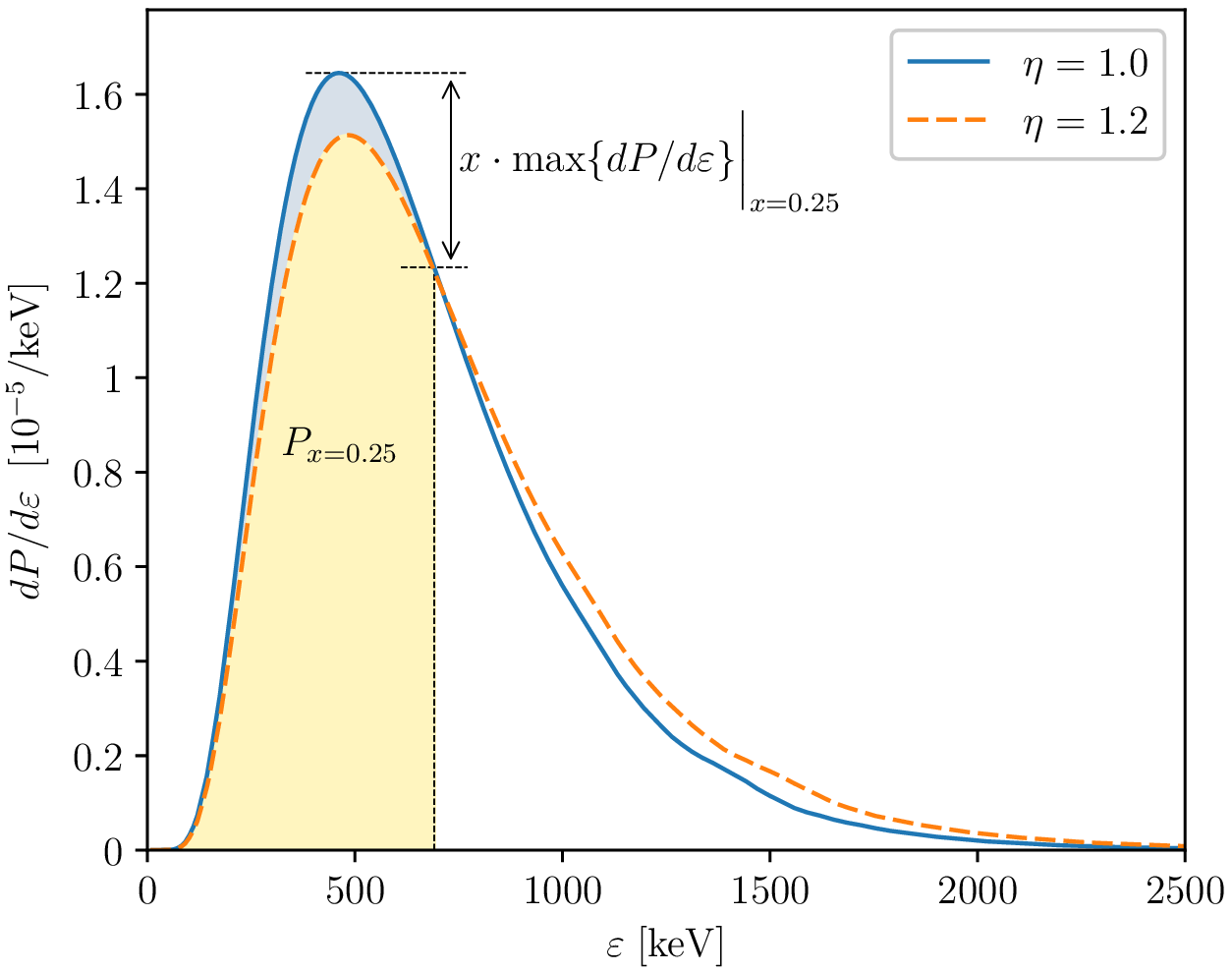}
\end{center}
\caption{The definition of the partial probability $P_x$ for $x=0.25$ in the case of
  the U$^{92+ }-$U$^{92+}$ collision. $P_x$  accounts only for the positrons with the energies
  not exceeding the abscissa
value corresponding to the point $dP/d\varepsilon=(1-x)(dP/d\varepsilon)_{\rm max}$
on the right-hand side from the positron-spectrum maximum at $\eta=1$ ($\theta=180^o$).
The area shaded yellow corresponds to $P_x(\eta=1.2)$, while  $P_x(\eta=1)$
  incorporates the areas shaded yellow and blue.
}
\label{fig:def_px}
\end{figure}

\begin{figure}
\begin{center}
  \includegraphics[width=\columnwidth]{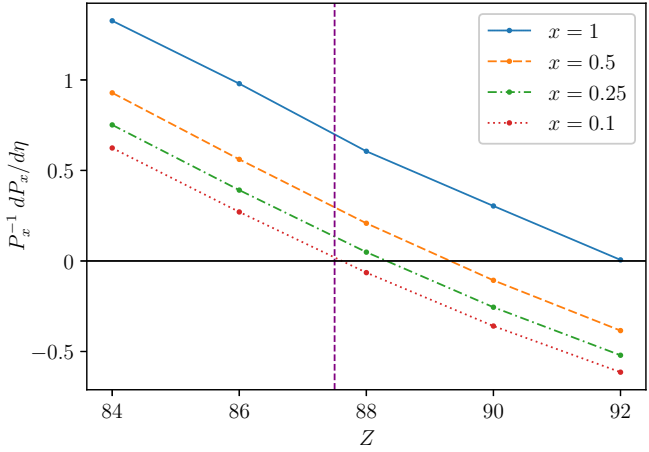}
\end{center}
\caption{The derivative $d\log{P_x}/d\eta =(1/P_x)dP_x/d\eta$
at $\eta=1$ ($\theta=180^o$)
  for different $x$ in the symmetric ($Z=Z_1=Z_2$) collisions.
The dashed vertical line at $Z \approx 87.5$ indicates the boundary
between the subcritical and the supercritical regime.
}
\label{fig:lnP}
\end{figure}

\begin{figure}
\begin{center}
  \includegraphics[width=\columnwidth]{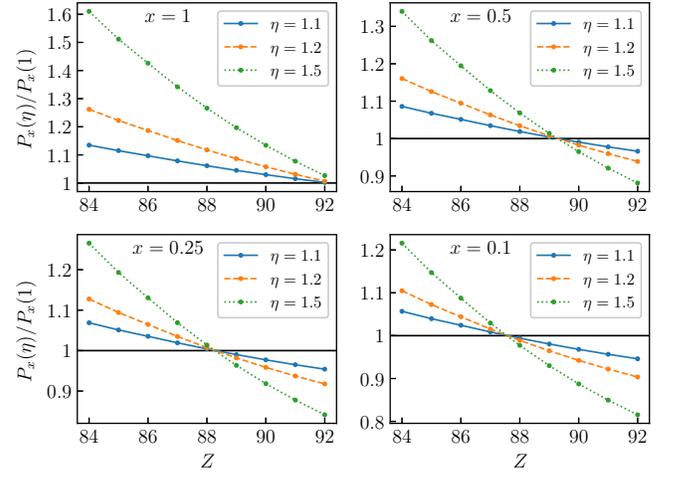}
\end{center}
\caption{The ratio of $P_x$ evaluated at different $\eta$ to its value
at $\eta=1$ ($\theta=180^o$) as a function of $Z=Z_1=Z_2$.
The transition from the region $P_x(\eta)/P_x(1)>1$ to the region $P_x(\eta)/P_x(1)<1$
indicates the transition from the subcritical to the supercritical mode.
}
\label{fig:pxpx}
\end{figure}

\begin{figure}
\begin{center}
\includegraphics[width=\columnwidth]{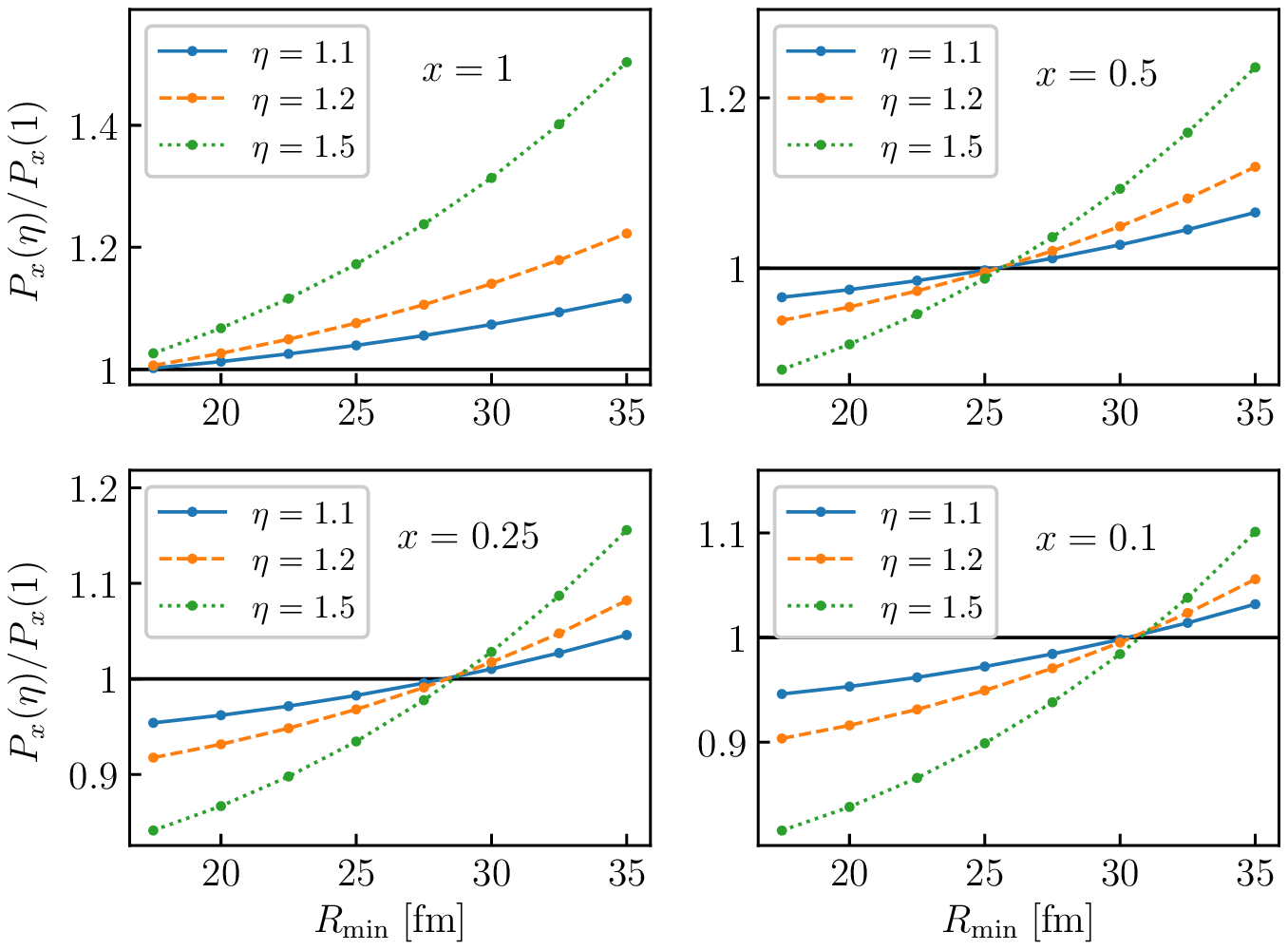}
\end{center}
\caption{The ratio of $P_x$ evaluated at different $\eta$ to its value
  at $\eta=1$ ($\theta=180^o$)
  as a function of $R_{\rm min}$ for the U$^{92+}-$U$^{92+}$ collision.
The transition from the region $P_x(\eta)/P_x(1)>1$ to the region $P_x(\eta)/P_x(1)<1$
indicates the transition from the subcritical to the supercritical mode ($R_{\rm cr}=32.7$ fm).
}
\label{fig:pxpx92}
\end{figure}

As one can see from Figs.~\ref{fig:sp1},~\ref{fig:sp2},~\ref{fig:sp3},
and \ref{fig:spectrum92}, the changes in the positron spectra caused
by the transition to the supercritical mode are mainly concentrated
at the energies around the spectrum maximum and lower,
where the spontaneous pair creation takes place.
This fact can be used to quantify the signature
of the transition to the supercritical mode in terms
an integral characteristic of the positron spectrum
which accounts only for a part of it instead of the total
probability. It is natural to restrict this part
to the region, where the changes of interest take place.
To this end, we introduce
a partial probability $P_x$, which is defined as shown in Fig.~\ref{fig:def_px}.
This means that the probability $P_x$ accounts only for the positrons
with the energies not exceeding the abscissa
value corresponding to the point $dP/d\varepsilon=(1-x)(dP/d\varepsilon)_{\rm max}$
on the right-hand side from the positron-spectrum maximum at $\eta=1$ ($\theta=180^o$).
The parameter $x$ should be chosen to cover the area which includes the positrons
with the energies within the range $\Gamma_{\rm dyn}$ from the resonance energy.
As discussed above,
in the case of the  U$^{92+}-$U$^{92+}$ collision
the energies of  the spontaneous positrons are mainly limited by about
600 keV.  It follows that the parameter $x$ should be chosen not less than
$x\approx 0.1$, which
 corresponds to the inclusion of about 50\% of the total probability.

In Fig.~\ref{fig:lnP} we display the   derivative $d\log{P_x}/d\eta =(1/P_x)dP_x/d\eta$
 at $\eta=1$ ($\theta=180^o$)
for
$x=0.1, 0.25, 0.5, 1.0$ in the symmetric ($Z=Z_1=Z_2$) collisions
in the range $Z=84-92$. As one can see from
the figure, the use of  $d\log{P_x}/d\eta|_{\eta=1}$ at $x=0.1, 0.25$, and even at $x=0.5$ (instead
of $x=1$  corresponding to the total probability)
leads to a strong increase of the effect we are interested in. Indeed, in the case of the U$^{92+}-$U$^{92+}$
collision,
the derivative  $d\log{P_{x}}/d\eta|_{\eta=1}$
amounts to about $-0.6$ at  $x=0.1$, $-0.5$ at $x=0.25$,  and $-0.4$ at $x=0.5$,
while for the total probability ($d\log{P_{x=1}}/d\eta|_{\eta=1}$) it is almost zero. This means
 taking into account only a part of the positron spectrum, as defined above,
allows to get a much stronger evidence of the transition to the supercritical mode.
This statement is supported by Fig.~\ref{fig:pxpx}, where
we present the ratio of $P_x$ evaluated at different $\eta$ to its value
at $\eta=1$ ($\theta=180^o$) as a function of $Z=Z_1=Z_2$.
The transition from the region $P_x(\eta)/P_x(1)>1$ to the region $P_x(\eta)/P_x(1)<1$
indicates the transition from the subcritical to the supercritical mode.
It can be seen that for $x=0.1, 0.25, 0.5$ this transition
takes place at lower $Z$ values than for $x=1$ (total probability).
In Fig.~\ref{fig:pxpx92}, we present the same ratio, $P_x(\eta)/P_x(1)$,
for the  U$^{92+}-$U$^{92+}$ collision as a function of $R_{\rm min}$.
Again, this figure clearly confirms the advantage of studying a part of the spectrum
compared to the total one and makes the uranium-uranium collisions
very promising for observing the vacuum decay.

\begin{figure}
\begin{center}
  \includegraphics[width=\columnwidth]{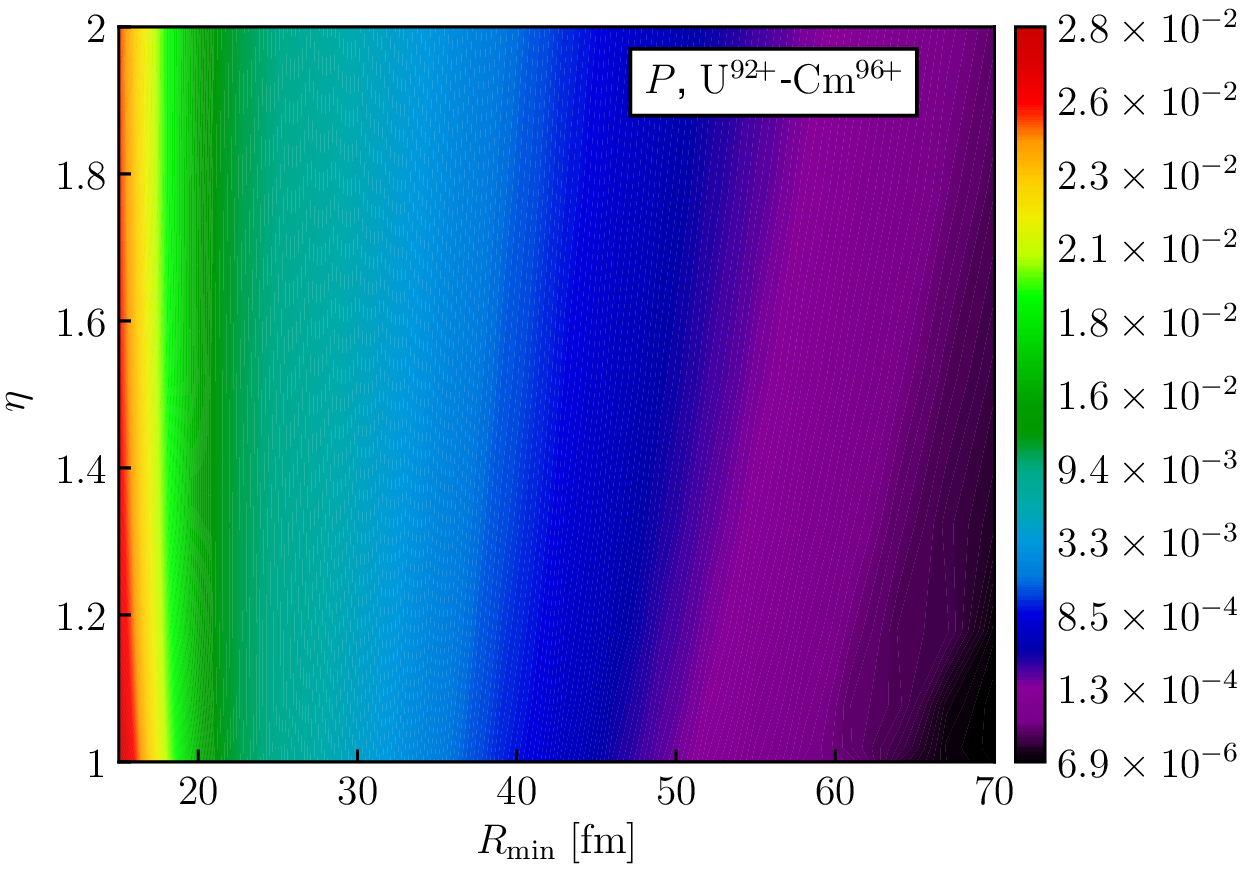}
\end{center}
\caption{The pair-creation probability in non-symmetric U$^{92+}$ $-$ Cm$^{96+}$
  collisions
    as a function of the minimal internuclear distance $R_{\rm min}$ and
the parameter $\eta=E/E_0$, which is the ratio of the collision energy $E$
to the energy of the head-on collision at the same $R_{\rm min}$.
}
\label{fig:P_rmin_92-96}
\end{figure}

\begin{figure}
\begin{center}
  \includegraphics[width=\columnwidth]{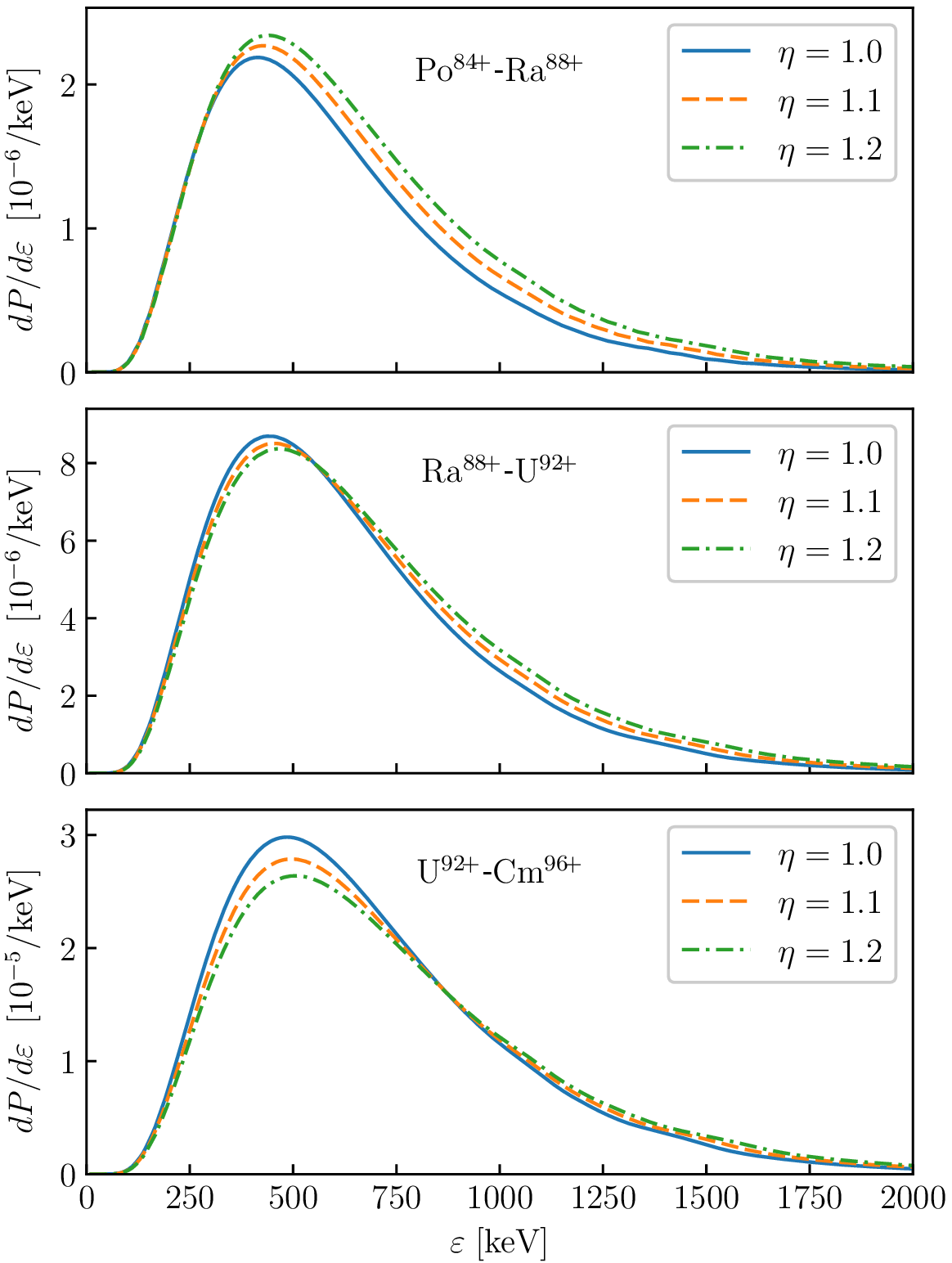}
\end{center}
\caption{The positron spectra
 for non-symmetric
 Po$^{84+}$ $-$ Ra$^{88+}$
 Ra$^{88+}$ $-$ U$^{92+}$, and
 U$^{92+}$ $-$ Cm$^{96+}$ collisions
at $R_{\rm min}=17.5$ fm and  $\eta=E/E_0=1.0, 1.1, 1.2$.}
\label{fig:sp4}
\end{figure}

\subsection{Pair-creation probabilities and positron spectra for non-symmetric collisions}

All the above studies can be directly extended to non-symmetric collisions.
In Fig.~\ref{fig:P_rmin_92-96} we present  the pair-creation probability in
non-symmetric  U$^{92+}$ $-$ Cm$^{96+}$ collisions   as a function of the minimal
internuclear distance $R_{\rm min}$ and
the parameter $\eta=E/E_0$.
As in the case of the symmetric Cm$^{96+}$ $-$ Cm$^{96+}$
collisions (see Fig.~\ref{fig:P_rmin_96}),
the pair-creation probability
$P$ starts to increase with decreasing  $\eta$ when the system enters deeply enough
into the supercritical regime, which means that $R_{\rm min}$ becomes significantly
smaller than the critical value, $R_{\rm cr}\approx 40$ fm.
In contrast, in the subcritical region
$P$  decreases with decreasing $\eta$
(at a given value of  $R_{\rm min}$).
 In Fig.~\ref{fig:sp4}
 we present the positron spectra for non-symmetric
 Po$^{84+}$ $-$ Ra$^{88+}$,
 Ra$^{88+}$ $-$ U$^{92+}$, and
 U$^{92+}$ $-$ Cm$^{96+}$ collisions.
As in the case of the symmetric collisions, the behavior
of the positron spectra indicates a clear signature of the transition to the supercritical
regime.

\subsection{Pair creation in collisions of bare nuclei with neutral atoms}

In the previous sections, we studied the pair creation in collisions of bare nuclei only.
In this section, we will consider whether it is possible in principle to extend these studies
to collisions of bare nuclei with neutral atoms. While the direct calculation
of the pair production in such collisions requires development of special
methods that are beyond the scope of this work,
some rough estimates can be made by  calculating
the probability of a vacancy in the $1\sigma$ state of a quasimolecule
formed in such a collision. This is due to the fact that,
in accordance with the results for collisions of bare nuclei,
the main contribution to the pair creation is given
by the process with the electron captured into the lowest-energy bound state \cite{mul88,mal17,pop18,mal18}.
Therefore, the pair-creation probability is mainly defined by
availability of a vacancy in the $1\sigma$
level when the nuclei approach each other.
In this section,  we
evaluate the probability of a vacancy in the $1\sigma$ state at the critical
distance
for the  U$-$U$^{92+}$ and   Cm$-$U$^{92+}$ collisions
using the two-center method developed in Ref. \cite{tup12}.

\begin{figure}
\begin{center}
  \includegraphics[width=\columnwidth]{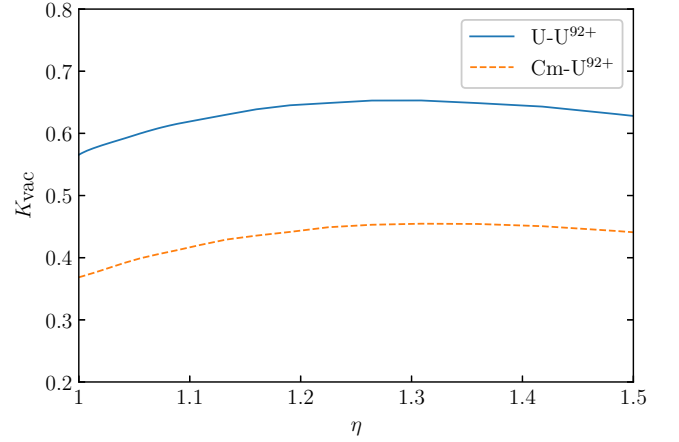}
\end{center}
\caption{ The coefficient $K_{\rm vac}$ evaluated by Eq. (\ref{Kvac})
 for the  $1\sigma$ level
at the critical internuclear distances, 27.0 fm and 34.8 fm for the  U$-$U$^{92+}$ and
Cm$-$U$^{92+}$ collisions, respectively,
as a function of the collision energy, $\eta=E/E_0$,
where $E_0$ is the head-on collision energy.
The calculations are performed
for the Rutherford trajectories with $R_{\rm min}=17.5$ fm.
}
\label{fig:K_vac}
\end{figure}

The quantum dynamics of electrons in the
collision of the neutral U and Cm atoms with the bare U nucleus
is described by the time-dependent Dirac equation with the two-center
potential given by a sum of the potential of the neutral atom and
the Coulomb potential of the bare nucleus.
The potential of the neutral atom
is defined by the density functional theory (DFT) in the local density approximation
(LDA) using the Perdew-Zunger (PZ) parametrization~\cite{Perdew_81} (see Ref. \cite{tup12}
for details).
The basis is formed by the Dirac-Fock (DF) orbitals for the occupied states
and by the Dirac-Fock-Sturm (DFS) orbitals for the vacant (virtual) states,
localized at each of the two centers (see  Refs.  \cite{tup10,tup12} and references
therein). The DF and DFS orbitals are obtained in the
central field approximation by numerical solution of
the related integro-differential DF and DFS equations.
The radial DF and DFS
Hamiltonians describe the atom (ion) in the external field induced by
the other ion (atom). This external field is defined by the DFT
method in the LDA approximation with the PZ parametrization.
To the leading order,
the external potential effect can be taken into account by including the
external potential within the monopole approximation, when
only the spherically-symmetric part of the reexpansion of
the potential of the other atom (ion) is taken into account.
It should be noted that the basis constructed in this way
changes as the internuclear distance changes.
In other words, this basis depends on the time and the corresponding
time derivatives must be added to the Hamiltonian matrix \cite{tup10}.
The basis expansion coefficients of a set of one electron time-dependent
wave functions $\psi_i (t)$ are defined using the direct evolution
(exponential) operator $\hat U$ method \cite{tup10}. These coefficients
are represented by the columns of the $U$ evolution matrix
(the matrix of the operator $\hat U$).

The probability of pair creation with electron captured
into a vacancy at the quasi-molecular $1\sigma$ level
can be roughly estimated as
\begin{equation}
P(C)=2 \, P(C/\bar A \, B) \cdot K_{\rm vac} \,,
\end{equation}
where $P(C/\bar A \, B)$ is the conditional probability of
a pair being created and the vacancy being filled,
provided that there is exactly one $1\sigma_{1/2}$
vacancy (event $\bar A$) and the state $1\sigma_{-1/2}$ is occupied (event B).
The coefficient $K_{\rm vac}$ takes into account the probability of filling
one or two vacancies and the factor $2$ accounts for two values of the angular
momentum projection $\mu=\pm 1/2$ (see the Appendix for details).

To calculate the probability of a vacancy in the
$1\sigma$ level of the Hamiltonian of a quasimolecule at a given time
we use the approach which is described in the Appendix.
This approach is equivalent to a method which was previously used
for the evaluation of inclusive probabilities
(see Refs. \cite{Ludde_1985,Kirchner_2004} and references therein).

We define the time-dependent one-particle density operator $\hat \rho(t)$
and one-hole density operator  $\hat {\bar\rho}(t)$
\begin{eqnarray}
\hat \rho(t)=
\sum_i q_i \mid \psi_i(t) \rangle \, \langle \psi_i(t) \mid \,, \qquad
\hat {\bar\rho}(t)= \hat I - \hat \rho(t)\,,
\end{eqnarray}
where $q_i=0,1$ is the number of electrons on a spin-orbital $\psi_i$, which
is defined by the initial condition.
The inclusive probability $P_n$ of finding $n$ states occupied in an $N$-electron
system while the remaining $N - n$ electrons are not detected
is given by the determinant of an $n \times n$ matrix $\gamma$ constructed from
the one-electron density matrix \cite{Ludde_1985}
\begin{eqnarray}
P_{n}= \det{\gamma} \,,  \qquad
\gamma_{kj}= \langle \varphi_k \mid \hat \rho \mid \varphi_j \rangle\,,
\end{eqnarray}
where $\varphi_k$ are the stationary wavefunctions of the Hamiltonian at a given time,
and $k,j=1,\ldots , n$.
Similarly, the probability $\bar P_n$  of finding $n$ states unoccupied
is determined by
\begin{equation}
\bar P_{n}= \det{\bar \gamma} \,, \qquad
\bar \gamma_{kj}= \delta_{kj} - \gamma_{kj} \,.
\end{equation}
In this paper, we investigated the probability $\bar P_n$ of finding two
vacancies ($n=2$).
As a result, the coefficient $K_{\rm vac }$ has been
calculated using the following expression (see the Appendix)
\begin{equation} \label{Kvac}
K_{\rm vac }=1-\gamma_{11}\,.
\end{equation}
As the occupied $q_i=1$ spin orbitals $\psi_i(t)$, we consider all single-electron states
of the ground-state configuration of the neutral atom U or Cm and all states of the
negative-energy Dirac continuum.
All the positive-energy states of the bare uranium nucleus are considered as unoccupied.
The occupation of single-electron states of the negative-energy spectrum
corresponds to the Dirac picture and takes into account the Pauli exclusion
principle and pair production processes.
By means of the equation (\ref{Kvac}) we calculated
the $1\sigma$ vacancy probabilities
for the Rutherford trajectories presented in  Fig.~\ref{fig:tr} ($R_{\rm min}=17.5$ fm).
In Fig.~\ref{fig:K_vac} we present the coefficient $K_{\rm vac }$
at the critical internuclear distance,
which equals to 27.0 fm and 34.8 fm for the  U$-$U$^{92+}$ and
Cm$-$U$^{92+}$ collisions, respectively,
as a function of the collision energy, $\eta=E/E_0$,
where $E_0$ is the head-on collision energy.
Assuming $K_{\rm vac}$ remains of the same order of magnitude
 throughout the  supercritical region as at its boundary
(see Fig.~\ref{fig:K_vac}), we believe that the pair-creation probability
should be only a few times less, compared to the collisions of bare nuclei.
Therefore, due to relatively large values
of $K_{\rm vac}$,
all scenarios considered above for the collisions of bare nuclei
can be applied to the collisions of bare nuclei with neutral atoms,
provided the corresponding calculations of quantum dynamics of electrons
are performed.
In particular, the very convincing contrast in the transition from the subcritical
to the supercritical Cm-U system, which was observed for the positron
spectra in collisions of
bare nuclei  (see Fig.~\ref{fig:sp4}),
makes the Cm$-$U$^{92+}$ collision very promising for further
studies.

\section{Conclusion}
We have studied possible scenarios to access QED in
 the supercritical Coulomb field which can be created by heavy
 nuclei in low-energy collisions near the Coulomb barrier.
 This has been done by the calculations of  pair production
  in the collisions along
the trajectories, which correspond to a given minimum
 internuclear distance.  The clear signatures
 indicating the transition from the subcritical
 to the supercritical regime
 have been found in both  pair-creation
 probabilities and  positron spectra.
 The evidence of such a transition becomes even more convincing if
 one considers
 only a part of the positron spectrum around its maximum.
 For instance, it is very well pronounced in collisions
 of two uranium nuclei.
 In contrast to the
 previous studies by other authors, the scenarios considered
 do not require any sticking of colliding nuclei.
Low-energy collisions of a bare uranium nucleus
with neutral uranium and curium atoms have been also studied and
the probability of a vacancy in the lowest
energy level of a quasimolecule formed in such collisions has been evaluated.
These calculations showed that the same scenario can be applied to
access QED at the supercritical field in collisions of  bare nuclei with neutral atoms.
In particular,
examinations of the changes in the positron spectra when switching
from subcritical to supercritical Cm$-$U$^{92+}$ collisions are
very promising for this purpose.
The experimental study of the proposed scenarios, which seems
 feasible with the future facilities at  GSI/FAIR \cite{gum09,les16,hagXX},
 HIAF \cite{ma17}, and
 NICA \cite{ter15},
 would either prove the vacuum decay in the supercritical Coulomb field
 or lead to discovery of a new physics, which is beyond the presently
 used QED formalism.


 $$
 \,
 $$
 \section*{Acknowledgments}

We thank I. B. Khriplovich and Yu. Ts. Oganessian for stimulating discussions.
This work was supported by RFBR-Rosatom (Grant No. 20-21-00098),
by RFBR (Grants No. 18-03-01220, No. 20-02-00199, and No. 18-32-20063), and by the President
of the Russian Federation (Grant No. MK-1626.2020.2).
The work of R.V.P., V.M.S., and I.A.M. was also supported by
the Foundation for the advancement of theoretical physics
and mathematics ``BASIS''.  V.M.S. also acknowledges the support of the CAS
President International Fellowship Initiative (PIFI) and of SPbSU (COLLAB 2019: No 37722582).
Y.S.K. acknowledges the support from the CAS PIFI
under Grant No. 2018VMC0010. The work of R.V.P., N.V.K., and  D.A.T.
was also supported by TU Dresden via the DAAD Programm Ostpartnerschaften.
A.I.B. acknowledges the support from the Ministry of Science and
Higher Education of the Russian Federation (Grant No. 0784-2020-0025).
The research was carried out using computational resources provided by the Resource Center
``Computer Center of SPbSU''.

%
%
$$
\,
$$

\section*{Appendix: Probability of finding  core holes}
\setcounter{equation}{0}
\renewcommand {\theequation}{A\arabic{equation}}

\subsection {Reduced density matrix approach to a system of N-independent electrons}
The reduced density matrix (RDM) of $n$-th order for an $N$-electron system is defined by
\cite{McWeeny_1992, Davidson_1976}
\begin{widetext}
\begin{equation}
\begin{array}{lll} \displaystyle
\rho_n(x_1, \ldots ,x_n, x_1^{\prime}, \ldots , x_n^{\prime})
& = & \displaystyle
\binom{N}{n} \,
\int dx_{n+1}, \ldots, dx_N \, \Psi(x_1, \ldots, x_n, x_{n+1}, \ldots, x_N) \,
\\[6mm] &\times& \displaystyle
\Psi^{\ast}(x_1^{\prime},\ldots, x_n^{\prime},x_{n+1}, \ldots, x_N) \,,
\end{array}
\label{dens1}
\end{equation}
\end{widetext}
where $x$ stands for both position (${\bf r}$) and bispinor ($\tau=1,\ldots,4$) variables,
the integration over
$x$ implies the integration over ${\bf r}$ and the summation over $\tau$, and
$\Psi(x_1, \ldots, x_N)$ is the wave function of the system.
We can also consider the density operator $\hat \rho_n$ as a
linear integral operator with the density matrix $\rho_n$ being its kernel.
Then, the probability $P_n$ of finding $n$-electron system in
a state $\Phi_n$ is given by \cite{Davidson_1976}
\begin{equation}
P_n = n! \, \langle \Phi_n \mid \hat \rho_n \mid \Phi_n \rangle \,.
\label{prob1}
\end{equation}
Consider a system of independent $N$ electrons, described by a
set of one-electron wave functions $\psi_i$ ($i=1,\ldots ,N)$.
The $n$-th order RDM in this system is given
by \cite{Davidson_1976,McWeeny_1992}
\begin{widetext}
\begin{equation}
  \rho_n(x_1,\ldots x_n; x_1^{\prime}, \ldots ,x_n^{\prime})
  =\frac{1}{n!} \, \left |
\begin{array}{cccc} \displaystyle
\rho_1(x_1,x_1^{\prime}) & \rho_1(x_1,x_2^{\prime}) & \ldots &
\rho_1(x_1,x_n^{\prime})
\\[2mm] \displaystyle
\rho_1(x_2,x_1^{\prime}) & \rho_1(x_2,x_2^{\prime}) & \ldots &
\rho_1(x_2,x_n^{\prime})
\\[2mm] \displaystyle
\ldots      & \ldots      & \ldots & \ldots      \\[2mm] \displaystyle
\rho_1(x_n,x_1^{\prime}) & \rho_1(x_n,x_2^{\prime}) & \ldots &
\rho_1(x_n,x_n^{\prime})
\end{array} \right | ,
\end{equation}
\end{widetext}
where $\rho_1(x,x^{\prime})$ is the one-particle density matrix defined by
\begin{equation}
\rho_1(x,x^{\prime}) = \sum_{i=1}^N \psi_i(x) \, \psi_i^{\ast}(x^{\prime}) \,.
\end{equation}
The wave function $\Phi_n$ of an $n$-electron state is given by the Slater
determinant constructed from one-electron functions $\varphi_k$. In this case it is easy to
show that the probability $P_n$ defined by Eq. (\ref{prob1}) is equal to the
determinant of an $n \times n$ matrix $\gamma$ constructed from the
one-electron density matrix,
\begin{equation}
P_n = \det \gamma \,, \qquad
\gamma_{kj} = \langle \varphi_k \mid \hat \rho_1 \mid \varphi_j \rangle\,,
\label{prob2}
\end{equation}
where $\hat \rho_1$ is the one-particle density operator defined by
\begin{equation}
\hat \rho_1 = \sum_{i \in \rm occupied}
\mid \psi_i \rangle \, \langle \psi_i \mid \,.
\end{equation}
Thus, $P_n$ is the probability of finding $n$ states
occupied in the $N$-electron system while the remaining $N - n$ electrons
are not detected. The equation (\ref{prob2}) coincides with that obtained
by the method of inclusive probabilities (see Refs. \cite{Ludde_1985, Kirchner_2004}
and references therein).

We can also introduce the probability $\bar P_m$ of finding $m$
holes in an $N$-electron system by defining the one-hole density
matrix $\bar \rho_1$. Due to completeness of the set of functions
$\psi_i(x)$ the hole density operator can be written in the form
\begin{equation}
\hat {\bar \rho}_1 =
\sum_{i \in \rm vacant} \mid \psi_i \rangle \,
\langle \psi_i \mid = \hat I -  \hat \rho_1 \,.
\end{equation}
Then, we get
\begin{equation}
\bar P_m=\det \bar \gamma \,, \qquad
\bar \gamma_{kj} = \langle \varphi_k \mid \hat {\bar \rho}_1 \mid
\varphi_j \rangle = \delta_{kj} -  \gamma_{kj} \,.
\label{prob3}
\end{equation}
%
\subsection {Probability of finding one or two core holes ($m=2$)
}
%
Consider an $N$-electron system and two one-electron wave functions $\varphi_1(x)$
and  $\varphi_2(x)$ which correspond to $1\sigma$ states with the
angular momentum projection $\mu=1/2$ and $\mu=-1/2$, respectively.
We denote by $A$ the event when
$1\sigma_{1/2}$ is occupied by electron and by $\bar A$ the event when
this state is vacant. Similarly, the event $B$ means that the state $1\sigma_{-1/2}$ is occupied
and the event $\bar B$ means that  $1\sigma_{-1/2}$ is  vacant.

Thus, four incompatible events can occur:  $A \, B$,
$A \, \bar B$, $\bar A \, B$ and $\bar A  \, \bar B$,
which correspond to the cases: both states are occupied,
the first state is occupied and the second state is vacant,
the first state is vacant and the second state is occupied,
and both states are vacant. Using the equations (\ref{prob2}) and (\ref{prob3}),
we obtain for the corresponding probabilities
\begin{equation}
\begin{array}{lll} \displaystyle
P(A\, B) &=& \displaystyle \gamma_{11} \, \gamma_{22} - |\gamma_{12}|^2\,,
\\[4mm]  \displaystyle
P(\bar {A} \, \bar {B}) &=& \displaystyle
(1-\gamma_{11}) \, (1-\gamma_{22}) - |\gamma_{12}|^2\,.
\end{array}
\end{equation}
Using $P(\bar A\, B)=P(\bar A) - P(\bar A \, \bar B)$, we obtain
\begin{equation}
\begin{array}{lll} \displaystyle
P(A\, \bar B) &=& \displaystyle \gamma_{11} \, (1-\gamma_{22})+|\gamma_{12}|^2\,,
\\[4mm]  \displaystyle
P(\bar {A} \, B) &=& \displaystyle  (1-\gamma_{11}) \, \gamma_{22}+
|\gamma_{12}|^2\,.
\end{array}
\end{equation}
With the help of the total probability rule,  for the probability
of pair creation with the occupation of at least one $1\sigma$ vacancy, $P(C)$, we obtain
\begin{equation}
\begin{array}{lll} \displaystyle
P(C) &=& \displaystyle
P(C/A \, B) \cdot P(A \, B)+
P(C/\, \bar A \, B) \cdot P( \, \bar A \, B)+
\\[5mm]  &+& \displaystyle
P(C/A \, \bar B) \cdot P(A \, \bar B)+
P(C/\bar A \, \bar B) \cdot P(\bar A \, \bar B)\,.
\end{array}
\end{equation}
Here $P(C/D)$ is the conditional probability, i.e. the probability of
occurring the event $C$, given the event $D$ has occurred.
Assuming the pair-creation probability is small enough,
we obtain
\begin{eqnarray}
P(C/\bar A \, \bar B) \simeq \, P(C/\bar A \, B)+ P(C/A \,\bar B) \,.
\end{eqnarray}
Then, taking into account that $P(C/A \, B)=0$, we get
\begin{equation}
P(C) \simeq
P(C/\bar A \,B) \left[1-\gamma_{11}\right ]+
P(C/A \, \bar B) \left[1-\gamma_{22}\right ]\,.
\end{equation}
Assuming the states with $\mu=\pm 1/2$ are equivalent, which means
$$
\gamma_{11}=\gamma_{22} \,, \qquad P(C/\bar A \, B)= P(C/A \, \bar B) \,,
$$
we have
\begin{equation}
P(C) \simeq 2 \, P(C/\bar A \, B) \, K_{\rm vac} \,,
\end{equation}
where
\begin{equation}
K_{\rm vac}=P(\bar A)=1-\gamma_{11} \,.
\end{equation}




\end{document}